\documentclass[11pt,aps,preprint,nofootinbib,superscriptaddress,nobalancelastpage]{revtex4}

\pdfoutput=1
\usepackage{graphicx}
\usepackage{amsmath}
\usepackage{amssymb}
\usepackage{float}
\usepackage{color}

\usepackage{hyperref}
\hypersetup{colorlinks,citecolor=blue,linkcolor=black,urlcolor=blue}

\def\bea{\begin{eqnarray}}
\def\eea{\end{eqnarray}}
\def\ba{\begin{eqnarray}}
\def\ea{\end{eqnarray}}
\def\be{\begin{equation}}
\def\ee{\end{equation}}
\def\beq{\begin{equation}}
\def\eeq{\end{equation}}

\newcommand{\lsim}{\mathrel{\rlap{\lower4pt\hbox{\hskip1pt$\sim$}}
    \raise1pt\hbox{$<$}}}         
\newcommand{\gsim}{\mathrel{\rlap{\lower4pt\hbox{\hskip1pt$\sim$}}
    \raise1pt\hbox{$>$}}}         

\newcommand{\leftrightarrowraised}{\mathrel{\rlap{\lower-0pt\hbox{\hskip1pt$\partial$}}
    \raise6 pt\hbox{$\leftrightarrow$}}}

\def\pM{\mathrel{\lower 1.25pt \hbox{\tiny(}\! 
                 \raise 1pt \hbox{$\, + \,$}
                 \settowidth {\dimen03} {$\, +\, $}
                 \hskip-\dimen03
                 \raise -2.4pt \hbox {$\, - \,$}
                 \!\lower 1.5pt \hbox{\tiny)}}}

\def\kth{Department of Physics,
School of Engineering Sciences, KTH Royal Institute of Technology,
AlbaNova University Center, 106 91 Stockholm, Sweden}

\def\ift{Instituto de F\'isica Te\'orica UAM/CSIC,
Calle Nicol\'as Cabrera 13-15, Cantoblanco E-28049 Madrid, Spain}

\def\coepp{
ARC Center of Excellence for Particle Physics at the Terascale (CoEPP), University of Adelaide, Adelaide, SA 5005, Australia}

\begin{document}

\preprint{IFT-UAM/CSIC-18-019}
\preprint{ADP-18-4/T1052}

\title{The distribution of inelastic dark matter in the Sun}

\author{Mattias Blennow}
\email{m.blennow@csic.es}
\affiliation{\kth}
\affiliation{\ift}
\author{Stefan Clementz}
\email{scl@kth.se}
\affiliation{\kth}
\author{Juan Herrero-Garcia}
\email{juan.herrero-garcia@adelaide.edu.au}
\affiliation{\coepp}
\begin{abstract}
If dark matter is composed of new particles, these may become captured after scattering with nuclei in the Sun, thermalise through additional scattering, and finally annihilate into neutrinos that can be detected on Earth. If dark matter scatters inelastically into a slightly heavier ($\mathcal{O} (10-100)$ keV) state it is unclear whether thermalisation occurs. One issue is that up-scattering from the lower mass state may be kinematically forbidden, at which point the thermalisation process effectively stops. A larger evaporation rate is also expected due to down-scattering. In this work, we perform a numerical simulation of the capture and thermalisation process in order to study the evolution of the dark matter distribution. We then calculate and compare the annihilation rate with that of the often assumed Maxwell--Boltzmann distribution. We also check if equilibrium between capture and annihilation is reached and find that this assumption definitely breaks down in a part of the explored parameter space. We also find that evaporation induced by down-scattering is not effective in reducing the total dark matter abundance.
\end{abstract}

\maketitle

\section{Introduction}

A popular class of models for explaining a number of observations in astrophysical systems is that of particle dark matter (DM)~\cite{Bergstrom:2000pn,Bertone:2004pz,Feng:2010gw,Bertone:2016nfn}. In the event that DM interacts with the particles of the standard model (SM), many different methods for observing it have been proposed. Over the last decades a number of experiments have been able to place impressive bounds on the DM mass and its interaction cross sections.

One of the many ways of searching for DM is to look for the effects that it may have on the Sun as it is captured by scattering against solar material~\cite{Press:1985ug,Gould:1987ir}. If DM annihilates (e.g., thermal relics), SM particles can be produced, which in turn decay or otherwise interact to give rise to a flux of high energy neutrinos or, in more exotic scenarios, to other SM particles. The neutrinos produced can be searched for in neutrino telescopes on Earth~\cite{Silk:1985ax,Krauss:1985aaa,Hagelin:1986gv,Gaisser:1986ha,Srednicki:1986vj,Griest:1986yu}, with various collaborations having performed such searches with no positive detection~\cite{Avrorin:2014swy,Choi:2015ara,Adrian-Martinez:2016gti,Aartsen:2016zhm}. The accumulation of large amounts of DM in the Sun may also affect helioseismology and the solar temperature. These modifications can potentially lead to observational effects with the possibility to constrain DM properties or alleviate the solar composition problem~\cite{Spergel:1984re,Lopes:2001ra,Bottino:2002pd,Frandsen:2010yj,Cumberbatch:2010hh,Taoso:2010tg,Lopes:2012af,Lopes:2013xua,Lopes:2014aoa,Vincent:2014jia,Vincent:2015gqa,Blennow:2015xha,Chen:2015poa,Vincent:2016dcp,Geytenbeek:2016nfg,Busoni:2017mhe}.

A collection of DM models that could possibly be probed by the production of neutrinos from DM annihilations inside the Sun is that of inelastic DM~\cite{TuckerSmith:2001hy}. These models were originally introduced to reconcile the annual modulation observation of the DAMA/LIBRA experiment~\cite{Bernabei:2010mq} with the null results of the CDMS experiment~\cite{Abusaidi:2000wg}, which ruled out its explanation in terms of a standard elastically scattering DM particle. Now DAMA is also incompatible with a number of other experiments, including the large Xenon-based experiments LUX~\cite{Akerib:2016vxi}, PandaX~\cite{Cui:2017nnn}, and XENON1T~\cite{Aprile:2017iyp}. It should be mentioned that it is also difficult to reconcile DAMA with other direct detection experiments in the case of inelastic scattering~\cite{Bozorgnia:2013hsa}. These models are also tightly constrained by direct detection (DD) experiments~\cite{Akimov:2010vk,Ahmed:2010hw,Aprile:2011ts,Chen:2017cqc}.

The evasion of the bounds from CDMS in inelastic DM models came from the introduction of a small mass splitting $\delta$ that separated two different DM states which, upon scattering, change from one to the other. In the scattering process, some energy is converted from kinetic energy to mass, which gives this type of model its name. The introduced mass splitting has a large impact on the scattering kinematics of DM and translates into altered solar capture rates of inelastic DM. Capture of inelastic DM in the Sun has been discussed for both the cases of endothermic and exothermic scattering in, e.g., Refs.~\cite{Nussinov:2009ft,Menon:2009qj,Shu:2010ta,McCullough:2013jma,Blennow:2015hzp,Smolinsky:2017fvb}. See also Refs.~\cite{McCullough:2010ai,Hooper:2010es,Baryakhtar:2017dbj} for studies of inelastic DM capture by compact stars such as white dwarves and neutron stars. Inelastic DM has also been proposed as a solution to the small scale structure problems in models where a light mediator induces large self-scattering cross sections~\cite{Schutz:2014nka,Zhang:2016dck,Blennow:2016gde}, and in models where the more massive state is unstable~\cite{SanchezSalcedo:2003pb,Abdelqader:2008wa,Bell:2010qt}.

Upon being captured inside the Sun, DM is often assumed to ``instantaneously" thermalise with the surrounding plasma, in which case its number density distribution is well described by a Boltzmann distribution with a specific temperature, i.e., $f \sim {\rm exp}(-E/T)$, at all times, see e.g.~Refs.~\cite{Griest:1986yu,Zentner:2009is}. In Ref.~\cite{Widmark:2017yvd} it has also been shown using numerical simulations that the Boltzmann distribution is a reasonable assumption and that the thermalised DM is generally concentrated in the core of the Sun. Therefore significant annihilation can occur. However, in the case of inelastic DM with sizable $\delta$, scattering of a particle in the lower mass state is only kinetically allowed if a large amount of kinetic energy is supplied to the collision. This implies that the DM particles scatter only off high velocity (and thus Boltzmann suppressed) solar nuclei, so that the scattering probability is very small, or that their orbit takes them from a radius with a large gravitational potential into regions closer to the solar center, which provides the necessary kinetic energy. Moreover, when a DM particle subsequently scatters from the higher to the lower mass state it may be boosted by the significant amount of energy that is released to a velocity such that it is no longer gravitationally bound. With this in mind, it is not obvious that a thermalised distribution is to be expected in the case of inelastic DM, nor is it clear if and when evaporation has to be taken into account. Both of these phenomena can have an impact on the annihilation rate of captured DM. A non-thermal distribution could alter the annihilation rate so that a larger population of DM must be present in order for equilibrium between solar capture and annihilation to occur, while evaporation would reduce the total number of particles that can annihilate. The assumption of capture-annihilation equilibrium allows one to bypass the annihilation rate in favour of a direct link between neutrino rates and the solar capture rate. Due to the effects of inelastic scattering on the annihilation rate it is unclear if this is a justified assumption.

In this paper, we study the thermalisation process of inelastic DM using numerical simulations. We analyse the impact that inelastic scattering has on the annihilation and evaporation rates. In Sec.~\ref{sec:inel_DM} we discuss the inelastic DM framework and the relevant kinematic effects introduced by a mass splitting. In Sec.~\ref{sec:cap_scatter}, we describe our approach to the problem and the numerical implementation of the simulation. The results are presented and discussed in Sec.~\ref{sec:results}. Finally, we summarise and give our conclusions in Sec.~\ref{sec:conclusions}.

\section{Inelastic dark matter} \label{sec:inel_DM}

There are various scenarios in which inelastic DM appears naturally~\cite{Hall:1997ah,Cui:2009xq,Chang:2010en}. In the simplest models inelastic DM consists of two states $\chi$ and $\chi^*$, with masses $m_\chi < m_{\chi^*}$ that satisfy
\begin{equation}
m_{\chi^*} - m_\chi = \delta \, \ll m_\chi.
\end{equation}
Although the mass splitting is very small relative to the DM masses, it has a significant impact on the resulting differential rates in DD experiments, as was noted in Ref.~\cite{TuckerSmith:2001hy}. According to our definition of the DM masses, $\chi^*$ is the slightly more massive state, which indicates that the scattering process is endothermic when the incoming DM particle is a $\chi$, and exothermic when the incoming particle is a $\chi^*$. Below, we discuss the endothermic case. The case of exothermic scattering can be recovered by substituting $\delta \rightarrow -\delta$. We define up-scattering as the process in which a particle in the lower mass state scatters with the solar nuclei labelled by $A$ into the higher mass state, i.e., $\chi A \rightarrow \chi^*A$. Down-scattering refers to the opposite reaction, $\chi^* A \rightarrow \chi A$.

We primarily study inelastic DM that couples to protons and neutrons through spin-independent interactions. We disregard spin-dependent scattering as in the Sun it is mainly hydrogen that carries spin. The reason is that capture of the $\chi$ state requires $\delta$ to be so small that scattering is essentially elastic for DM masses heavier than a few times that of the proton, while $\chi^*$ can still be captured in significant numbers~\cite{Blennow:2015xha}. However, for the latter, upon being captured and as long as $\delta$ is non-negligible, these particles that are now in the $\chi$ state would find themselves unable to up-scatter. Over time, this would create a cloud of loosely bound DM particles that is far too diffuse for annihilation to take place efficiently. In other words, capture via spin-dependent interactions of inelastic dark matter is only interesting in the limit where $\delta$ is so tiny that the model is essentially elastic and thermalisation is expected. 

In the following we focus on the case where the galactic halo is composed of both $\chi$ and $\chi^*$ with equal abundances, in which case both species are captured by the Sun. This is plausibly the case, as the temperature at which the overall DM abundance freeze-out occurs is of the order $m_\chi/20$, which far exceeds the values of $\delta$ considered in this work.

\subsection{Scattering kinematics}

The scattering kinematics of DM is extensively covered in Ref.~\cite{Blennow:2015hzp}. Here we briefly review the scattering kinematics that are important for the discussions that follow.

When DM scatters inelastically, the only modification to the cross section with respect to the elastic case is a multiplying phase-space factor. If the elastic DM--nucleus scattering cross section is $\sigma_0$, then the inelastic scattering cross section $\sigma_{\rm inel}$ would be
\begin{equation} \label{eq:CS_relation}
\sigma_{\rm inel} = \sqrt{1-\frac{2\delta}{\mu_{\chi A} v_{\rm rel}^2}} \, \sigma_0 \, ,
\end{equation}
where $v_{\rm rel}$ is the relative speed between the DM particle and its target and $\mu_{\chi A}$ is the DM--target reduced mass. When DM scatters endothermically, the relative velocity between the DM particle and its target must exceed
\begin{equation} \label{eq:u_lower}
u_{\rm rel,lower} = \sqrt{\frac{2 \delta}{ \mu_{\chi A}}}
\end{equation}
or there is simply not enough kinetic energy to produce the heavier state. On the other hand, there is no such constraint for exothermic scattering, which is always kinematically allowed.

When a collision occurs, the solution to the energy and momentum conservation equations yields the largest recoil energy of the target solar nucleus, $E_{\rm max}$, and the smallest one, $E_{\rm min}$:
\begin{equation} \label{eq:E_maxmin}
E_{\rm max \, (min)} = 2 \frac{\mu_{\chi A}^2}{m_\chi m_A}E_{\rm i,kin} \left( 1 \pM \sqrt{1-\frac{m_\chi}{\mu_{\chi A} E_{\rm i,kin}} \delta} \right) - \frac{\mu_{\chi A}}{m_A} \delta \, ,
\end{equation}
where $m_A$ is the mass of the target nucleus and $E_{\rm i,kin} = E_{\rm i} - \phi(r)$ is the kinetic energy of the incoming DM particle. The allowed range of recoil energies $E_R$ for a given $E_{\rm i,kin}$ is $E_{\rm min} \leq E_R \leq E_{\rm max}$.

In the rest frame of the target nucleus, the angle $\theta$ between the velocity vectors of the incoming and outgoing DM particle satisfies
\begin{equation} \label{eq:outgoing_angle}
{\rm cos}(\theta) = \frac{E_{\rm i,kin} + E_{\rm f,kin} - m_A E_{\rm R} / m_\chi}{2 \sqrt{E_{\rm i,kin}E_{\rm f,kin}}}\,,
\end{equation}
where $E_{\rm f,kin} = E_{\rm f} - \phi(r)$ is the kinetic energy of the outgoing DM particle. This relation is needed to determine the change of trajectory of a DM particle as it scatters.

\section{Solar capture, thermalisation, and annihilation} \label{sec:cap_scatter}

The method we employ to investigate the thermalisation of inelastic DM is similar to that of Refs.~\cite{Gould:1987ju,Liang:2016yjf}. This section presents the implementation of the numerical simulation.

\subsection{The phase space evolution of a captured population}

The effective classical Hamiltonian $\mathcal{H}$ describing a particle moving in a central potential $\phi(r)$ is given by
\begin{equation} \label{eq:Orbital_Hamiltonian}
\mathcal{H} = m_\chi E = \frac{1}{2}m_\chi \dot{r}^2 + \frac{m_\chi L^2}{2 r^2} + m_\chi \phi(r) \, \equiv \frac{1}{2}m_\chi \dot{r}^2 + m_\chi V_{\rm eff}(L,r) .
\end{equation}
It is very convenient, as we have done, to define the reduced energy $E$ as the energy divided by the DM mass, as well as the reduced angular momentum $\vec{L} = \vec{r} \times \vec{v}$, given the fact that the orbit of a particle in a central potential is independent of its mass. Since $E$ and $L=|\vec{L}|$ are conserved quantities, it is convenient to describe the DM particle orbit by these quantities rather than position and velocity.\footnote{We assume spherical symmetry, i.e., orbits in different planes are equivalent. We therefore only use $L$, the total magnitude of $\vec{L}$, and not its 3 components.} For a given angular momentum $L$, the smallest energy $E_{\rm min}(L)$ of a particle with a trajectory that intersects the Sun is given by
\begin{equation} \label{eq:EminL}
E_{\rm min}(L) = \min_{r \leq R_\odot} V_{\rm eff}(L,r) \, .
\end{equation}
Taking the above into consideration, we find it convenient to define the combination $\alpha = (E,L)$ as a label for the phase space position of a particular DM particle.

The time evolution of the total number of captured DM particles $N(t)$ follows the differential equation
\begin{equation}
\dot{N} = C_\odot - E_{\rm evap} - \Gamma_{\rm ann} \, ,
\end{equation}
where $\dot{N} \equiv dN/dt$, $C_\odot$ is the solar capture rate, $E_{\rm evap} \propto N$ is the evaporation rate, and $\Gamma_{\rm ann} \propto N^2$ is the rate at which particles are annihilated.\footnote{Note that in the literature the last term is often written as $2\Gamma_{\rm ann} N^2$. In this work we absorb the factor of 2 into $\Gamma_{\rm ann}$ and use a differently normalised number density distribution function.} If evaporation is neglected, the equilibrium solution ($\dot{N} = 0$) of the evolution equation reads
\begin{equation} \label{eq:cap_anni_eq}
C_\odot = \Gamma_{\rm ann} \,,
\end{equation}
which implies that there is equilibrium between capture and annihilations. In order to rigorously test this condition one must calculate the annihilation rate. However, this requires knowledge of how the DM is distributed in the Sun. In our simulations, the distribution of particles is discretised in $E$ and $L$ such that $f_\alpha$ describes the number of particles in a particular state $\alpha$. The evolution of the distribution is then governed by the equation
\begin{equation} \label{eq:governing_eq}
\dot{f}_\alpha = \sum_\beta \Sigma_{\alpha \beta} f_\beta + C_\alpha - f_\alpha \sum_\beta \Gamma_{\alpha \beta} f_\beta \, .
\end{equation}
Each element in $\vec{f}$ contains the total number of particles in state $\alpha$, while each element in $\vec{C}$ gives the capture rate into the corresponding state. The off-diagonal elements in~$\Sigma_{\alpha\beta}$ $(\alpha \neq \beta)$ give the rate with which particles in state $\beta$ scatter against solar nuclei and end up in the state $\alpha$. The diagonal entries in $\Sigma$ are negative and correspond to the rate at which particles scatter from the corresponding state to all other states, including evaporation, i.e., positive energy states where the DM particle escapes the Sun's gravitational well. Finally, $\Gamma_{\alpha \beta}$ gives the rate at which a particle in state $\alpha$ annihilates with a particle in state $\beta$.\footnote{In the case of DM self-capture due to self-interactions, see e.g., Ref.~\cite{Zentner:2009is}, $\Sigma_{\alpha \beta}$ would also incorporate that effect.}

We also need to know the fraction of the time that a particle in a given state $\alpha$ finds itself at a radius $r$ as it travels between the maximal radius, $r_+$, and the minimal radius, $r_-$, of the complete orbit. These are found by solving Eq.~\eqref{eq:Orbital_Hamiltonian} with the substitution $\dot{r} = 0$. The time it takes the particle to move between radius $r_1$ to radius $r_2$ can be found by isolating $\dot{r}$ in Eq.~\eqref{eq:Orbital_Hamiltonian}, which leads to
\begin{equation} \label{eq:orbital_time}
T(r_1,r_2) = \int dt = \int_{r_1}^{r_2} \frac{dr}{\dot r} = \int_{r_1}^{r_2} \frac{dr}{\sqrt{2(E-V_{\rm eff}(L,r))}} \, .
\end{equation}
Integrating from $r_1 = r_-$ to $r_2 = r_+$ gives the time that a particle needs to complete half an orbit.

\subsection{Solar capture}

The derivation of the solar capture rate of DM particles originating from the DM halo dates back several decades~\cite{Press:1985ug,Gould:1987ir}. In the standard calculation any information on the DM energy and angular momentum post scattering is discarded, since any particle with $E < 0$ is counted towards the total capture. However, here we are not only interested in the total capture rate but also in the distribution of the captured particles in $E$-$L$ space. We therefore give a short description of how we compute $C_\alpha$.

The solar capture rate of DM in differential form is given by~\cite{Gould:1987ir}
\begin{equation} \label{eq:solar_cap}
dC = \pi n_\chi \frac{f(u)}{u} \frac{d\sigma}{dE_R}(w,E_R) \, n_A(r) \, \frac{2}{\sqrt{1-(L/rw)^2}} \, dr \, du \, dE_R \, dL^2 \, .
\end{equation}
In the above, we have assumed that the target nuclei are stationary. Here $r$ is the radius at which scattering occurs, $n_A(r)$ is the local density of target particles of species $A$, $u$ is the DM velocity at a distance where the gravitational potential of the Sun is negligible, and $w=\sqrt{u^2 + v_{\rm esc}(r)^2}$ is the velocity of the particle at radius $r$. The local halo number density of DM and its speed distribution at the location of the Sun enter explicitly through $n_\chi$ and $f(u)$. Note that the velocity distribution $\tilde{f}(\vec{u})$ is normalised such that
\begin{equation}
\int \, \tilde{f}(\vec{u}) \, d^3u = \int f(u) \, du = 1,
\end{equation}
where
\begin{equation}
f(u) = \int \, \tilde{f}(u,\theta,\phi) \, u^2 \, {\rm sin}(\theta) \, d\theta \, d\phi \, .
\end{equation}
The quantity $L^2$ is the square of the reduced angular momentum of the incoming DM particle from the DM halo, and $d\sigma / dE_R(w,E_R)$ is the differential cross section~\cite{TuckerSmith:2001hy}
\begin{equation}
\frac{d \sigma}{dE_R } (w,E_R) = \frac{m_A A^2 \sigma_{\chi p}}{2 \mu_{\chi p}^2 w^2} |F(E_R)|^2\, .
\end{equation}
Here it has been assumed that the coupling between DM and nuclei is isospin conserving, leading to the $A^2$ enhancement of the cross section, where $A$ is the total number of nucleons.\footnote{For isospin violating DM, $A^2$ is instead replaced by the factor $(Z+(A-Z) f_n/f_p)^2$, where Z is the number of protons and $f_p$ ($f_n$) is the coupling of DM to protons (neutrons). Note that $f_p$ can be absorbed in the definition of $\sigma_{\chi p}$.} Furthermore, $\sigma_{\chi p}$ is the DM--proton cross section entering as $\sigma_0$ in Eq.~\eqref{eq:CS_relation}, $\mu_{\chi p}$ is the DM--proton reduced mass, and $w$ is the relative velocity between the DM and the nucleus. Interestingly, the phase-space factor relating the inelastic and elastic scattering cross sections in Eq.~\eqref{eq:CS_relation} is cancelled. The form factor $F(E_R)$ accounts for the decoherence in the DM--nucleus scattering process when the momentum transfer $q(E_R)$ is large. The latter is related to $E_R$ by $q=\sqrt{2 m_A E_R}$.

In order to calculate $C_\alpha$, the integrand of Eq.~\eqref{eq:solar_cap} is discretised over the region of relevant $r$, $u$, $E_R$ and $L$. The integration range in $r$ is $0 < r < R_\odot$, where $R_\odot$ is the solar radius, and that for $L^2$ is $0 < L^2 < r^2 w^2$. The limits in $u$ and $E_R$ are complicated and we refer to the discussion in Ref.~\cite{Blennow:2015hzp}. To ensure that the mesh is fine enough we calculate the total capture rate as $C_\odot = \sum_\alpha C_\alpha$ as well as by integrating Eq.~\eqref{eq:solar_cap} as done in Ref.~\cite{Blennow:2015hzp} (but using the Helm form factor given in eq.~\eqref{eq:Helm} rather than the very frequently used exponential form factor). We have verified that the two agree to better than $1\,\%$ accuracy. 

At each discretisation point, the incoming DM velocity vector $\vec{w}_{\rm i}$ can be reconstructed. When the DM particle scatters, its energy post-collision is known from
\begin{equation} \label{eq:Eout}
E_{\rm f} = E_{\rm i} - E_R - \delta/m_\chi,
\end{equation}
where $E_{\rm i \, (f)}$ is the energy of the incoming (outgoing) DM particle. The outgoing DM speed $w_{\rm f}$ is known from the equation above. The angle $\theta$ between the incoming and outgoing velocity vector is given by Eq.~\eqref{eq:outgoing_angle}. There is also an azimuthal angle $\varphi$ around $\vec{w}_{\rm i}$ at which the outgoing velocity vector lies that is randomly distributed in the interval 0 to 2$\pi$. In terms of these angles, the angular momentum of the outgoing DM particle $L_{\rm f}$ is given by
\begin{equation} \label{eq:Lout}
L_{\rm f}^2 = r^2 w_{\rm f}^2 \left[ 1 - \left( \sqrt{1-\left( \frac{L}{r w} \right)^2 } {\rm cos}(\theta) - \frac{L}{r w} {\rm sin}(\theta) {\rm cos}(\varphi) \right)^2 \right] \, .
\end{equation}
We use Monte-Carlo methods to find the probability distribution for a scattering DM particle at each discretisation point to end up in a state $\alpha$. Finally, $\vec{C}$ is found by summing over all discretised states weighted by their probability densities.

\subsection{Scattering among different states}

When DM particles have been captured, occasional scattering with solar nuclei takes a particle initially in the $\beta = (E_{\rm i}$, $L_{\rm i})$ state into the state $\alpha = (E_{\rm f}$, $L_{\rm f})$. The differential scattering rate at radius $r$ of a DM particle with velocity $\vec{w}(r)$, travelling through a gas of nuclei of element $A$ with velocity $\vec{v}$, number density $n_A(r)$, and velocity distribution of the nuclei $f_A(r,\vec{v})$, is given by~\cite{Gould:1987ju}
\begin{equation} \label{eq:scattering_rate}
dR(r) = \sigma \, n_A(r) \, f_A(r,\vec{v}) \, |\vec{w}(r)-\vec{v}| \, d^3v \, .
\end{equation}
The velocities of nuclei in the Sun follow the Boltzmann distribution
\begin{equation}
f_A(r,\vec{v}) = \left( \frac{m_A}{2 \pi T(r)} \right)^{3/2} {\rm exp}\left( -\frac{m_A \vec{v}^2}{2 T(r)}  \right) \, ,
\end{equation}
where $T(r)$ is the temperature of the solar plasma at radius $r$. The cross section $\sigma$ that enters is the integral over the differential cross section in the frame in which the nucleon is stationary.

In order to find $\Sigma_{\alpha \beta}$ we discretise the Sun into thin spherical shells with radii $r_i$. Under the assumption that DM particles complete many orbits between interactions, the rate at which particles in state $\beta$ scatter at radius $r_i$ and end up in state $\alpha$ is given by
\begin{equation}
\mathcal{R}_{\beta \rightarrow \alpha}(r_i) = R_\beta(r_i) \, \mathcal{T}_{\beta}(r_i) \, \mathcal{P}_{\beta \rightarrow \alpha}(r_i) \, .
\end{equation}
Here, $R_\beta(r_i)$ is the total scattering rate at radius $r_i$ and $\mathcal{T}_{\beta}(r_i)$ is the fraction of the orbital time that the particle spends inside the shell. The factor $\mathcal{P}_{\beta \rightarrow \alpha}(r_i)$ is the probability that the particle ended up in the particular state $\alpha$ after it scattered at $r_i$. Having calculated the above, the off-diagonal elements in the $\Sigma$ matrix are given by the sum of contributions from all shells that the particle passes through on its orbit
\begin{equation} \label{eq:sigma_ab}
\Sigma_{\alpha \beta} = \sum_i \mathcal{R}_{\beta \rightarrow \alpha}(r_i) \, .
\end{equation}
The diagonal elements of the $\Sigma$ matrix are given by the negative of the total scattering rate from state $\alpha$,
\begin{equation}
\Sigma_{\alpha \alpha} = - \sum_i R_{\alpha}(r_i) \, ,
\end{equation}
which also includes evaporation.

To find $\mathcal{P}_{\beta \rightarrow \alpha}(r)$, the DM velocity vector $\vec{w}_{\rm i}$ of a particle in state $\beta = (E_{\rm i},L_{\rm i})$ before scattering is required. In the Sun's rest frame, its magnitude is found to be
\begin{equation} \label{eq:energy_vel}
w_{\rm i}(r) = \sqrt{2E_{\rm i} - 2 \phi(r)} \, .
\end{equation}
We are free to choose a coordinate system in which $\vec{r} = (r,0,0)$. The angle between $\vec{r}$ and $\vec{w}_{\rm i}(r)$ is given by $\xi = {\rm sin}^{-1}(L_{\rm i} / r w_{\rm i}(r))$. Thus the DM velocity vector can be written as
\begin{equation}
\vec{w}_{\rm i}(r) = w_{\rm i}(r)\, ({\rm c}_\xi, {\rm s}_\xi,0) \, ,
\end{equation}
where we use ${\rm s}_x = \sin(x)$ and ${\rm c}_x =\cos(x)$. The velocity vector of the nucleus can be parametrised as
\begin{equation}
\vec{v} = v \,( {\rm c}_\xi {\rm c}_\eta - {\rm s}_\xi {\rm s}_\eta {\rm c}_{\varphi_1} , {\rm s}_\xi {\rm c}_\eta + {\rm c}_\xi {\rm s}_\eta {\rm c}_{\varphi_1} , {\rm s}_\eta {\rm s}_{\varphi_1}) \,
\end{equation}
in terms of two other angles $\eta$ and $\varphi_1$ which are uniformly distributed in the intervals $0 < \eta < \pi$ and $0 < \varphi_1 < 2 \pi$, respectively. We have chosen the nuclei and DM velocity vectors to be aligned if $\eta = 0$. 

If scattering is kinematically allowed, a Galilean transformation is made to the frame in which the nucleus is stationary and the DM velocity is $\vec{w}_{\rm i,sc} = \vec{w}_{\rm i}(r) - \vec{v}$. It is in this reference frame that the recoil energy $E_R$ is defined and its allowed range is in the interval $[E_{\rm min}, E_{\rm max}]$, given by Eq.~\eqref{eq:E_maxmin}. For a given recoil energy, the angle $\theta$ between the outgoing DM velocity $\vec{w}_{\rm f,sc}$ and $\vec{w}_{\rm i,sc}$ can be calculated using Eq.~\eqref{eq:outgoing_angle}. Transforming back to the Sun's rest frame, the DM velocity after scattering is
\begin{equation}
\vec{w}_{\rm f} = \mathcal{R}(\varphi_2) \vec{w}_{\rm f,sc} + \vec{v} \, ,
\end{equation}
where the operator $\mathcal{R}(\varphi_2)$ rotates $\vec{w}_{\rm f,sc}$ around $\vec{w}_{\rm i,sc}$ by the angle $\varphi_2$, which is uniformly distributed in the interval $0 < \varphi_2 < 2\pi$. The state in which the DM particle ends up in is given by
\begin{equation}
E_{\rm f} = E_{\rm i} + \frac{1}{2} \left( w_{\rm f}^2 - w_{\rm i}^2(r) \right), \qquad L_{\rm f}^2 = r^2 w_{\rm f}^2 - (\vec{r} \cdot \vec{w}_{\rm f})^2 \, .
\end{equation}
The fractional time spent in the shell with inner radius $r_{\rm inner}$ and outer radius $r_{\rm outer}$ is calculated as
\begin{equation}
\mathcal{T}_\beta(r_i) = \frac{T(r_{\rm inner},r_{\rm outer})}{T(r_-,r_+)} \,,
\end{equation}
where $T(r_1,r_2)$ is given in Eq.~\eqref{eq:orbital_time}. The shell widths are chosen such that
\begin{equation}
r_{\rm inner} = \frac{r_{i-1}+r_i}{2}, \qquad r_{\rm outer} = \frac{r_i+r_{i+1}}{2} \, .
\end{equation}
The $\Sigma_{\alpha \beta}$ matrix is then found using Monte-Carlo methods.

\subsection{The radial number distribution function and the annihilation rate}

In order to calculate the annihilation rate of DM, detailed knowledge of the radial number density distribution function $f(r)$ is necessary. The annihilation rate for self-annihilating DM is given by
\begin{equation} \label{eq:anni_rate}
\Gamma_{\rm ann} = \, \int \, \sigma_{\rm ann}(\vec{v}_{\rm rel}) |\vec{v}_{\rm rel}| f(\vec{r},\vec{v}_1) f(\vec{r},\vec{v}_2) \, d^3v_1 \, d^3v_2 \, d^3r \, ,
\end{equation}
where $\vec{v}_{\rm rel} = \vec{v}_1-\vec{v}_2$ is the relative velocity between the two colliding particles, $f(\vec r,\vec{v})$ is the phase-space density distribution for DM, which has been normalised such that
\begin{equation}
\int f(\vec{r},\vec{v}) \, d^3v \, d^3r = N \, ,
\end{equation}
where $N$ is the total number of captured DM particles. We assume that the spatial distribution is spherically symmetric. However, a possible consequence of a very low number of scattering events of a particle over a long time may be a preference for some orbital planes over others due to a directional dependence of the flux of incoming DM particles. Such a directional dependence could be caused by, for example, the solar motion through the DM halo or anisotropies in the galactic distribution of DM. This would introduce an angular dependence in the number density distribution so that the local DM distribution is increased in some regions relative to the spherically symmetric case, which would increase the overall annihilation rate with respect to the latter case. In the following we assume that spatial spherical symmetry holds when evaluating Eq.~\eqref{eq:anni_rate}, keeping the above caveat at the back of the mind.

When the mean free path of DM inside the Sun is much larger than the solar radius (as considered here), the radial number density distribution is often approximated as an isothermal Maxwell--Boltzmann distribution that can be written as \cite{Spergel:1984re,Faulkner:1985rm,Griest:1986yu,Zentner:2009is,Blennow:2015xha,Garani:2017jcj}
\begin{equation} \label{eq:iso_dist}
f_{\rm iso}(\vec{r}) = n_0 \,{\rm exp}(-\vec{r}^{\,2}/r_\chi^2) \, N \, ,
\end{equation}
where $n_0 = \pi^{-3/2} \, r_\chi^{-3}$. Assuming constant solar density and temperature, the length scale $r_\chi$ of the distribution is given by
\begin{equation} \label{rchi}
r_\chi^2 = \frac{3 k_{B} T_c}{2 \pi G \rho_c m_\chi} \, ,
\end{equation}
where $k_{B}$ is the Boltzmann constant and $G$ is the gravitational constant. The bulk of the DM distribution is generally located in such a centralised region that the density~$\rho_c$ and temperature~$T_c$ can be approximated by the corresponding values at the Sun's center. For elastic DM, this assumption is generally valid for scattering cross sections that yield significant capture rates, see e.g.~Refs.~\cite{Nauenberg:1986em,Widmark:2017yvd}. In this case, the annihilation rate of DM, written in terms of the thermally averaged annihilation cross section $\langle \sigma_{\rm ann} v \rangle$, is
\begin{equation}\label{eq:anni_iso}
\Gamma_{\rm ann,iso} = \langle \sigma_{\rm ann} v \rangle \int f^2_{\rm iso}(\vec{r}) \, d^3r \, .
\end{equation}
It is clear that altering $f(\vec{r},\vec{v})$ could significantly modify the expected annihilation rate and that such a change is expected for inelastic DM if the sub-dominant elastic scattering is negligible. As our simulated distributions are given in $E$-$L$ space, we must map them onto $r$-$v$ space. With the assumption of many orbits per scattering, a DM particle in state $\alpha$ spends the fractional time $\mathcal{T}_\alpha(r)$ at radius $r$. The radial distribution function can thus be calculated by distributing all particles of each state into all possible radii, weighed by the fractional time spent at that radii:
\begin{equation} \label{eq:r_dist}
\tilde{f}_{\rm num}(r) = \sum_{\alpha} f_\alpha \mathcal{T}_\alpha(r) \, .
\end{equation}
This is just the angular averaged distribution of the full three-dimensional spatial distribution, i.e., 
\begin{equation}
\tilde{f}_{\rm num}(r) = r^2 \, \int \, f_{\rm num}(r,\theta,\varphi) \, {\rm sin}(\theta) \,  d\theta \, d\phi \, .
\end{equation}
Since spherical symmetry has been assumed, the relation above informs us that the three-dimensional distribution function $f(\vec{r})$ can be found from $\tilde{f}(r)$ as $f_{\rm num}(\vec{r}) = (4 \pi r^2)^{-1} \, \tilde{f}_{\rm num}(r)$. Unfortunately, solving Eq.~\eqref{eq:governing_eq} to find $f_\alpha$ can be computationally infeasible. Neglecting the annihilation rate, the analytic solution for $\vec{f}$ is
\begin{equation} \label{eq:dist_greens}
\vec{f}(t) = \int_0^t \, e^{\Sigma (t-t')} \, \vec{C}(t') \, dt' \, ,
\end{equation}
where a possible time-dependence in the solar capture rate has been taken into account. This allows us to find out if the total capture of DM is in equilibrium with the loss due to evaporation after a solar lifetime. It also permits us to calculate the annihilation rate, and compare it with that of an isothermal distribution.

The annihilation cross section times relative velocity can be expanded as
\begin{equation}
\sigma(v_{\rm rel}) v_{\rm rel} = a + b v_{\rm rel}^2 + \dots
\end{equation}
where $a$ is non-zero for s-wave annihilation. Below we make a simple comparison of s-wave annihilating DM from a Boltzmann distribution to the distribution that we extract from our numerical data. Since the relative velocities of DM particles in the Sun are small, if both $a, b\neq 0$, $a$ dominates and $\sigma(v_{\rm rel}) v_{\rm rel}$ is constant. In this case, we can trivially calculate the integrals over $\vec{v}_1$ and $\vec{v}_2$ in Eq.~\eqref{eq:anni_rate}. We then obtain the ratio between the s-wave annihilation rate for the derived distribution
\begin{equation} \label{eq:anni_calc}
\Gamma_{\rm ann,num}(t) = \langle \sigma_{\rm ann} v \rangle \int f^2_{\rm num}(\vec{r},t) \, d^3r \, ,
\end{equation}
and the isothermal distribution in Eq.~\eqref{eq:anni_iso}, i.e.,
\begin{equation} \label{eq:anni_comp}
\frac{\Gamma_{\rm ann,num}(t)}{\Gamma_{\rm ann,iso}} = \frac{\int f^2_{\rm num}(\vec{r},t) \, d^3 r}{\int f^2_{\rm iso}(\vec{r}) \, d^3 r} \, .
\end{equation}
This provides us with a quantitative measure of how much the annihilation rate is affected due to the change in the DM distribution relative to the Maxwell--Boltzmann distribution.

Note that if the annihilation cross section is velocity dependent the full phase-space distribution is required. Under the assumption of a spherical distribution, it can be found as follows. Any state that contributes to $\tilde{f}(r)$ at some radius $r$ gives a contribution to the velocity distribution at this radius. The magnitude of the velocity $v$ can be found from Eq.~\eqref{eq:energy_vel}, while the angle between the radial coordinate and the velocity vector is $\psi_1 = {\rm sin}^{-1}(L/r_i v)$. These two relations can be used to extract the two-dimensional velocity distribution $f(r_i,v,\psi_1)$. It should be recognised that, from the symmetry of the problem, we have that $f(r_i,v,\theta) = f(r_i,v,-\theta)$.

\section{Numerical results} \label{sec:results}

We must now make some assumptions in order to proceed. Specifically, we must define the galactic velocity distribution and the local background density of DM, as well as the nuclear form factor. We use the value $n_\chi = n_{\chi^*} = 0.2$~GeV/cm$^3$, which is half of the local DM density~\cite{Catena:2009mf,Read:2014qva,Pato:2015dua,Sivertsson:2017rkp}. In any example where elastic scattering is considered, we assign $n_\chi$ the value $0.4$~GeV/cm$^3$. We also assume the standard Maxwellian model for the galactic velocity distribution, with a shift to the solar frame,
\begin{equation}
f(u) = \frac{u}{\sqrt{\pi} v_{\odot}^2} \left[ {\rm exp}\left(-\frac{3}{2}\frac{(u-v_{\odot})^2}{\bar{v}^2}\right) - {\rm exp}\left(-\frac{3}{2}\frac{(u+v_{\odot})^2}{\bar{v}^2}\right) \right] \, .
\end{equation}
The solar velocity through the Milky Way, $v_\odot$, is taken to be 220~km/s, and the velocity dispersion $\bar{v} = 270$~km/s.

Unless otherwise stated, we use the DM-proton cross section $\sigma_{\chi p} = 10^{-42}$~cm$^2$. Both in the case of capture and subsequent scattering of captured particles, we use the Helm form factor~\cite{Helm:1956zz}
\begin{equation} \label{eq:Helm}
F(q) = 3 \frac{j_1(qR)}{qR} e^{-q^2s^2/2} \, ,
\end{equation}
where $q = \sqrt{2 m_A E_R}$ is the momentum transfer in the scattering process, $j_1$ is the spherical Bessel function of the first kind, and $R$ is given by
\begin{equation}
R = \sqrt{b^2+\frac{7}{3} \pi^2 a^2 - 5 s^2} \, .
\end{equation}
We use $a = 0.52$~fm, $s = 0.9$~fm, $b = (1.23 A^{1/3}-0.6)$~fm~\cite{Lewin:1995rx}. In order to speed up the computation time of $C_\alpha$ and $\Sigma_{\alpha \beta}$ we only take into account scattering on the elements hydrogen, helium, nitrogen, oxygen, neon, and iron, with radial abundances provided by the AGSS09ph solar model~\cite{Serenelli:2009yc}. This is an excellent approximation as the abundances of the other solar elements are negligible and contribute very little to scattering rates.

We use 100 individual states in $E$ that are uniformly distributed over all possible bound state energies. For every discretisation point in $E$, $L$ is uniformly discretised in 100 states between $0$ and $L_{\rm max}$, which is the largest allowed angular momentum for the given energy and can be found by inverting Eq.~\eqref{eq:EminL}. Therefore, in total, we use $10^4$ states.

The following plots are shown in units of energy ($E$) of $G M_\odot/R_\odot$, and in units of angular momentum ($L$) of $\sqrt{GM_\odot R_\odot}$, where $M_\odot$ is the solar mass. One can easily check that these quantities naturally correspond to the typical energy and angular momentum of a DM particle orbiting around the centre of the Sun:
\begin{align}
E_\chi &= m_\chi E = m_\chi\,\frac{G\, M_\odot}{R_\odot} \simeq 20\,\left( \frac{m_\chi}{10\,{\rm GeV}}\right)\, {\rm keV}\,,\\
L_\chi &=m_\chi L =  m_\chi\,\left(\frac{G\, M_\odot}{R_\odot}\right)^{1/2}R_\odot \simeq 0.03\,\left( \frac{m_\chi}{10\,{\rm GeV}}\right)\,{\rm GeV\,s}.
\end{align}
Furthermore, notice that $E_\chi \sim \delta$ for typical WIMP masses, and therefore it is expected that the excited state can be created by endothermic scatterings (see also the discussions in Refs.~\cite{TuckerSmith:2001hy,Nussinov:2009ft}).

\subsection{The distribution of captured particles}

\begin{figure}
\centering
\includegraphics[width=0.48\textwidth]{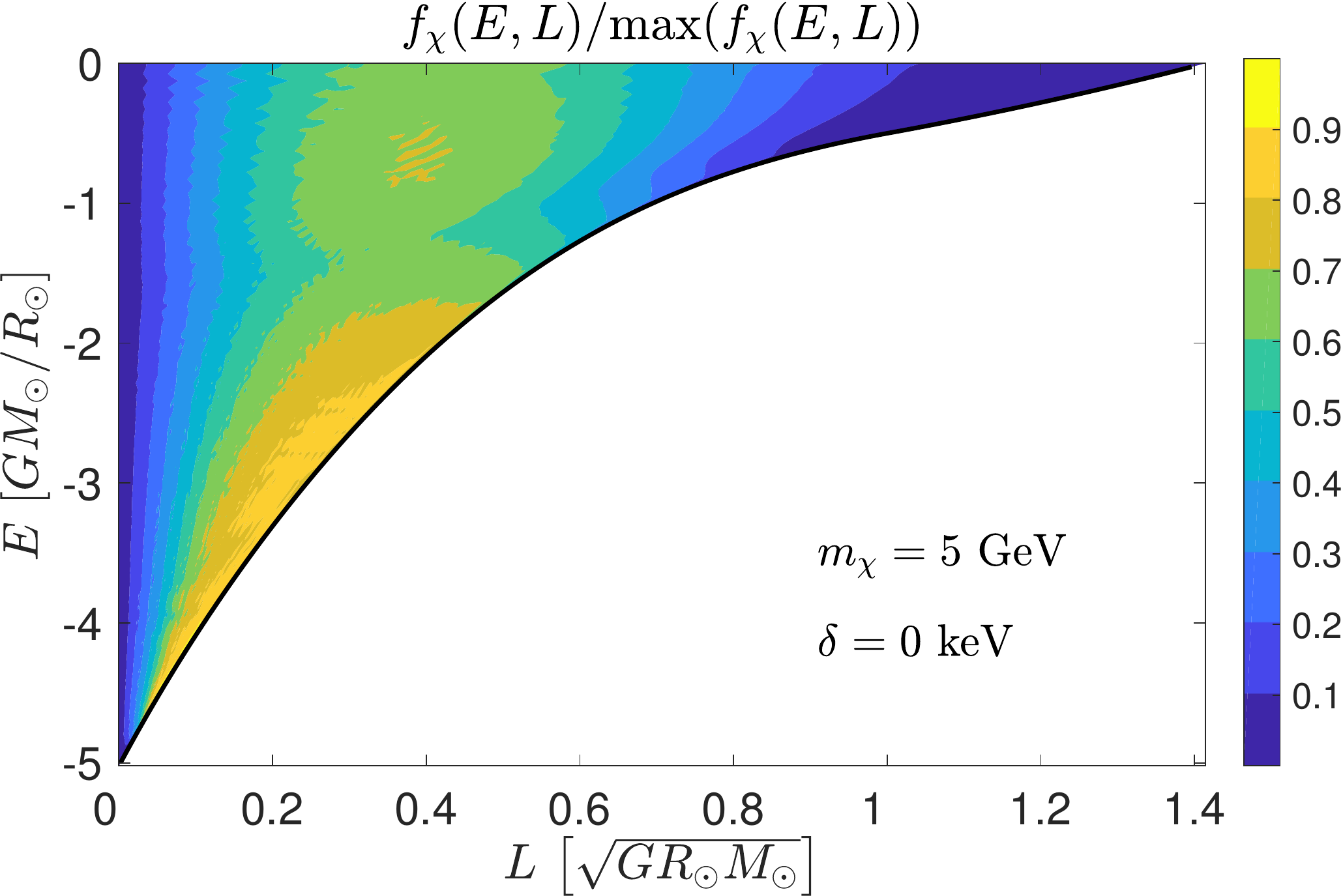}~~
\includegraphics[width=0.48\textwidth]{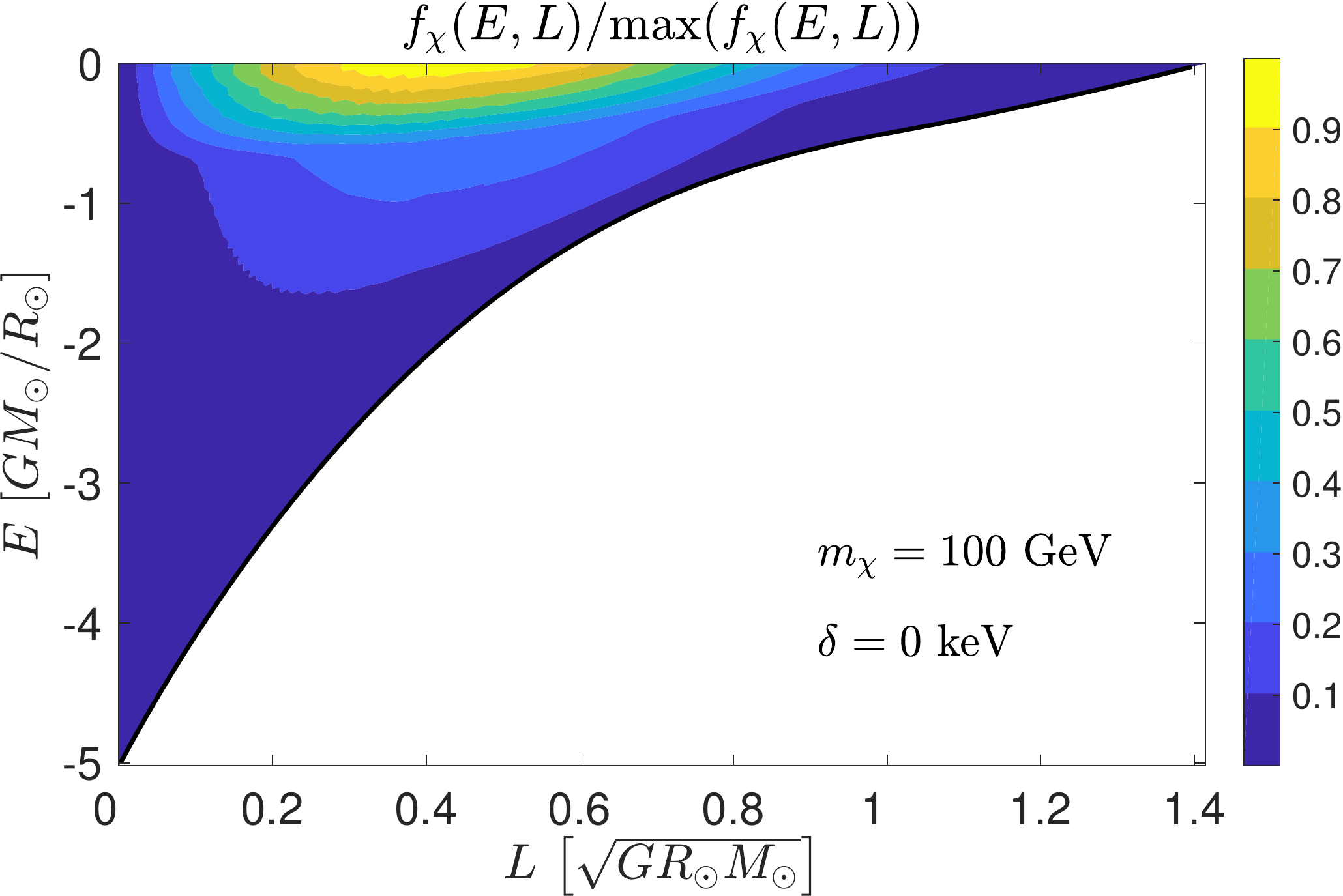}
\caption{\label{fig:mx5_caps} The density of dark matter capture in the $E$-$L$ plane normalised by the largest value of the distribution. Elastic scattering is assumed. \emph{Left panel:} $m_\chi = 5$~GeV. \emph{Right panel:} $m_\chi = 100$~GeV.}
\end{figure}

In Fig.~\ref{fig:mx5_caps}, we show the density of capture in $E$-$L$ space, normalised by its maximum value, for the elastic case, taking the DM mass $m_\chi = 5$~GeV in the left panel and $m_\chi = 100$~GeV in the right. For $m_\chi = 5$~GeV, capture is dominated by helium and oxygen, followed by a slightly lower capture rate by hydrogen and nitrogen. The concentration of capture in the region centred slightly above $E = -GM_\odot/R_\odot$ and $L \sim 0.3\sqrt{GM_\odot R_\odot}$ is due to scattering on hydrogen, which absorbs little recoil energy due to its low mass relative to the DM. For helium, oxygen and nitrogen, capture tends to be concentrated towards more strongly bound orbits, with a preference for more circular orbits, i.e., larger $L$. For $m_\chi = 100$~GeV, capture is primarily due to scattering on helium and oxygen in almost equal parts, at a rate that is a few times larger than that for capture by iron and neon. As can be seen, capture is now concentrated towards states that are much less bound. This is expected since the ability to lose energy in a collision for heavier DM is hampered by the relatively low mass of hydrogen and oxygen.

\begin{figure}
\begin{center}
\includegraphics[width=0.48\textwidth]{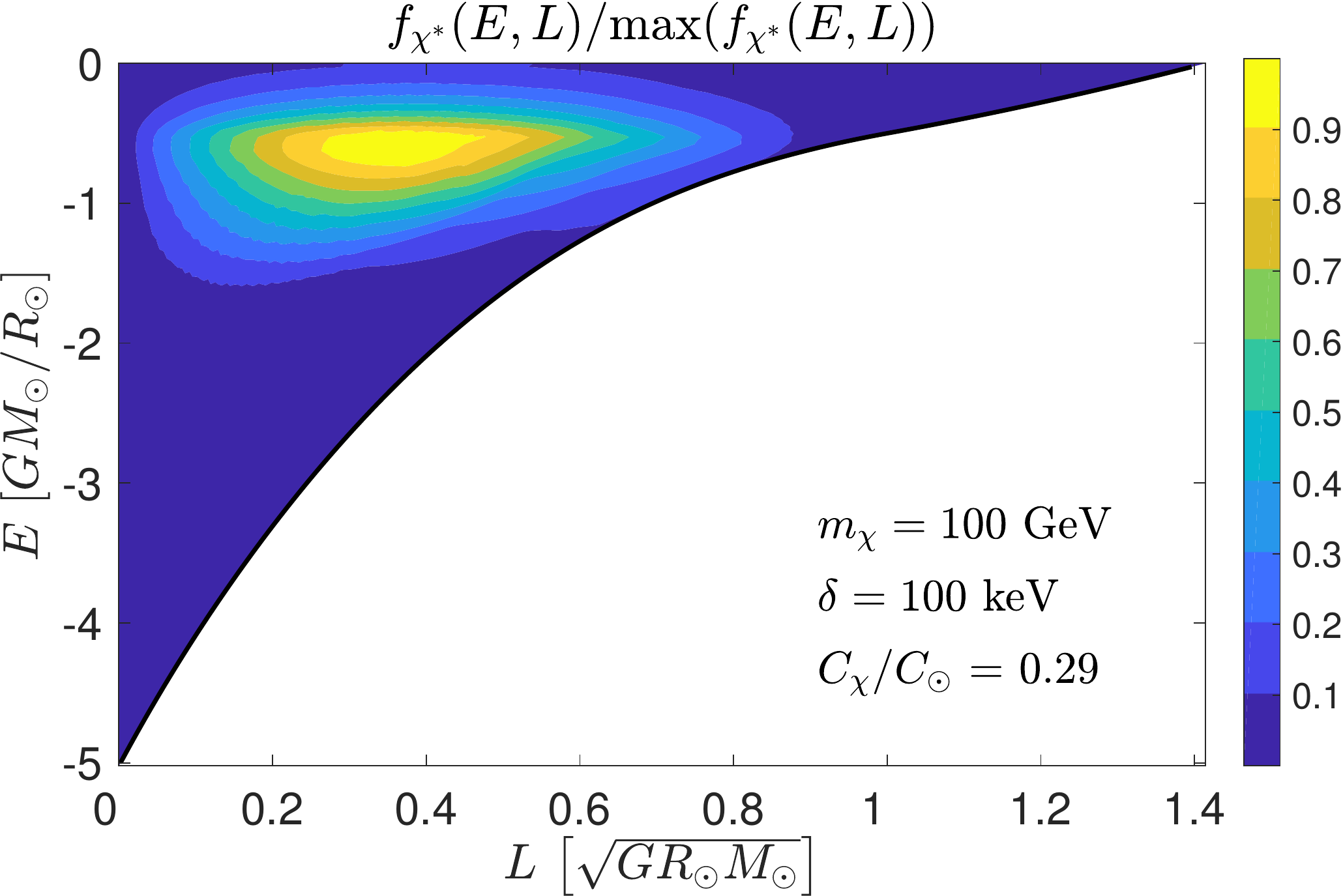}~~
\includegraphics[width=0.48\textwidth]{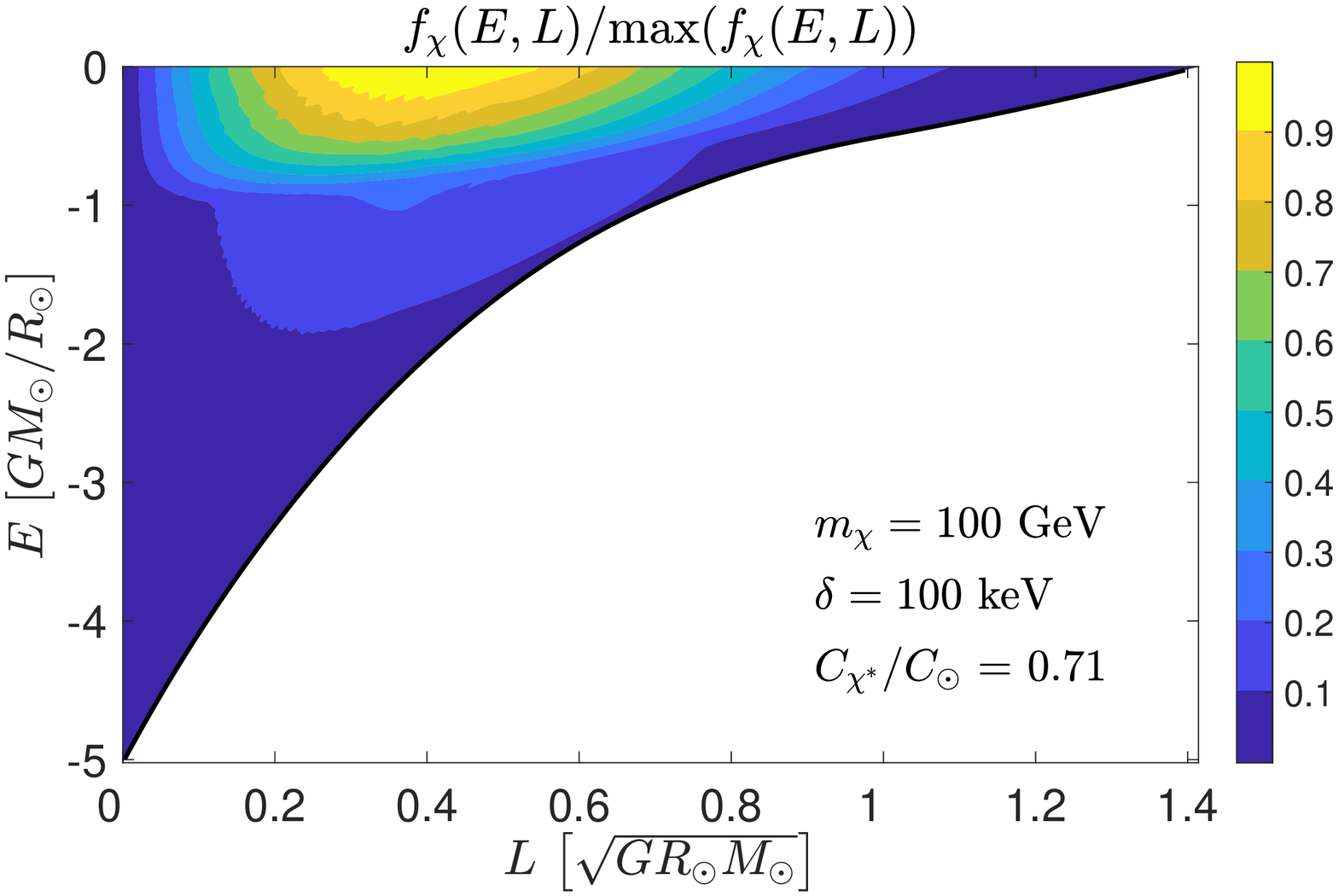}
\caption{\label{fig:mx100_caps} The density of dark matter capture  in the $E$-$L$ plane. Inelastic scattering is assumed, with $m_\chi = 100$~GeV and $\delta = 100$~keV. The distributions are normalised by their own largest value. \emph{Left panel:} Capture of halo $\chi$ particles.  \emph{Right panel:} Capture of halo $\chi^*$ particles.}
\end{center}
\end{figure}

Moving on to inelastic scattering, Fig.~\ref{fig:mx100_caps} shows an example of the density of capture of DM, with capture of $\chi$ in the left plot and $\chi^*$ in the right plot. We use $m_\chi = 100$~GeV and $\delta = 100$~keV. The capture rate of $\chi$ particles, $C_\chi$, is roughly half as large as the capture of~$\chi^*$, $C_{\chi^*}$, which is expected from previous studies, see~ e.g., Ref.~\cite{Blennow:2015hzp}. An interesting difference between the capture of $\chi$ and~$\chi^*$ particles is the fact that the former are captured into more tightly bound orbits than the latter. This is due to two reasons, the first of which is that a significant amount of kinetic energy is lost in the endothermic process to produce the~$\chi^*$. This loss of energy reduces the form-factor suppression as the momentum transfer is not as large. Scattering takes place primarily on iron, which due to its large mass is also a superb target for absorbing recoil energy relative to the other elements. On the other hand, energy being released in the exothermic case translates into a larger form factor suppression and thus a preference for scattering events in which the DM particle loses as little energy as possible, leaving it less tightly bound. In this case, capture occurs primarily due to scattering with helium nuclei and to a lesser degree with oxygen, both of which are not very efficient at absorbing recoil energy. Overall, the shape of the region into which capture proceeds through exothermic scatterings is similar to the elastic case with $m_\chi = 100$~GeV.

\subsection{Time evolution of the distribution in $E$-$L$ space}

\begin{figure} 
\centering
\includegraphics[width=0.48\textwidth]{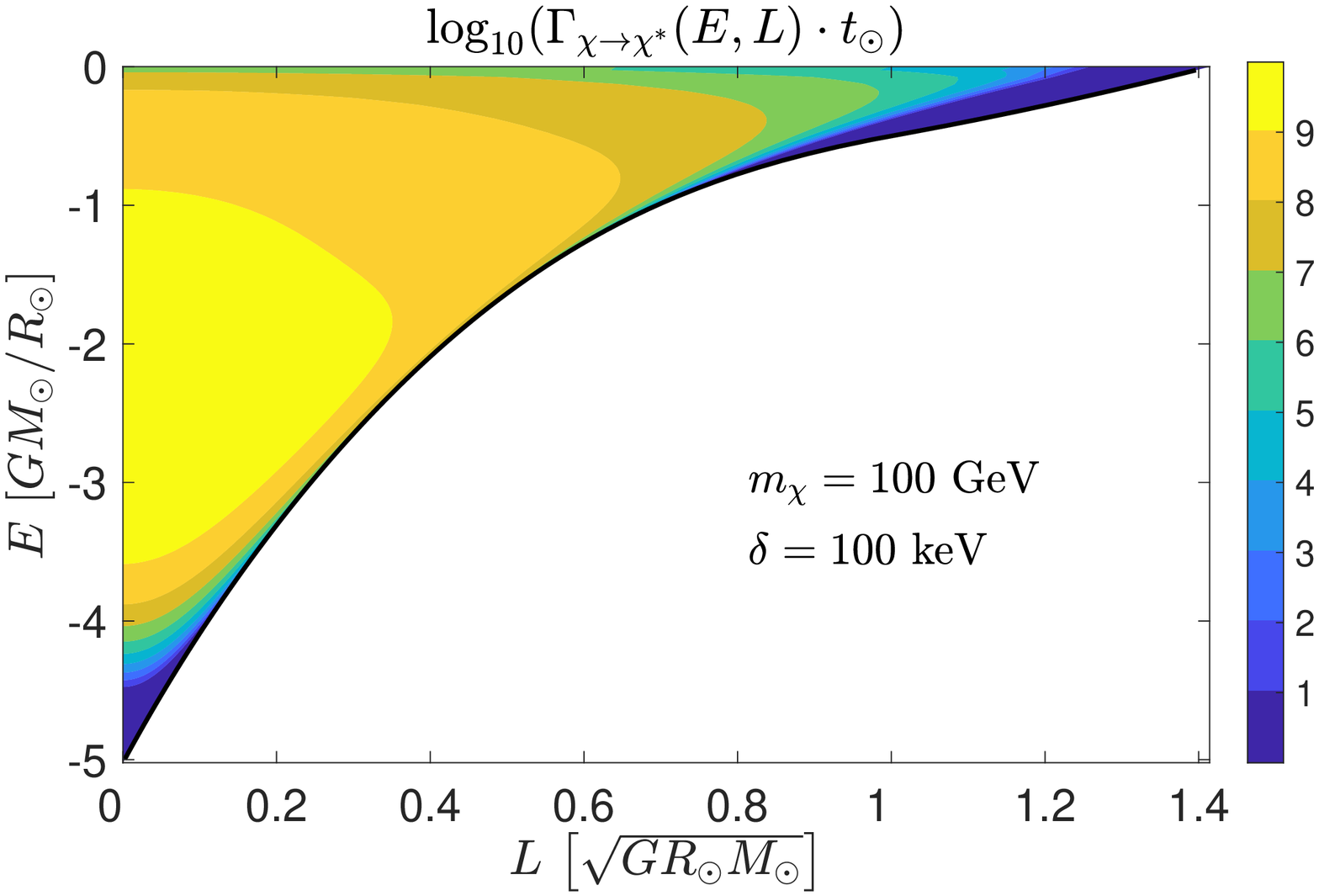}~~
\includegraphics[width=0.48\textwidth]{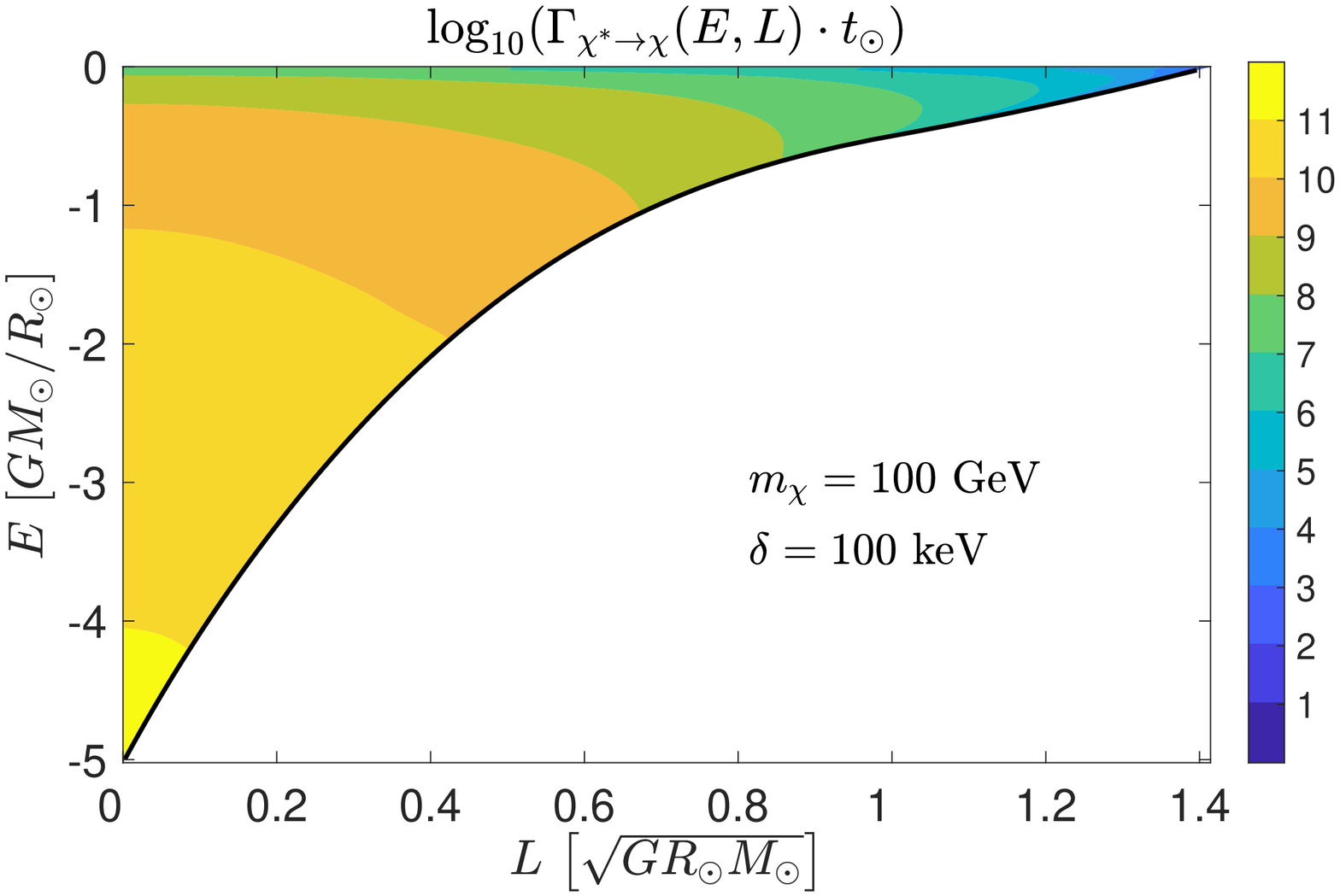}
\caption{The base 10 logarithm of the product of the total scattering rate and the solar lifetime  $t_\odot$ in the $E$-$L$ plane. We fix $m_\chi = 100$~GeV and $\delta = 100$~keV. \emph{Left panel:} Endothermic scattering. \emph{Right panel:} Exothermic scattering.} \label{fig:mx100_scatter} 
\end{figure}

Having calculated the $\Sigma_{\alpha \beta}$ matrix, the total scattering rate from each state can be found. The case with $m_\chi = 100$~GeV and $\delta = 100$~keV is shown in Fig.~\ref{fig:mx100_scatter}, where we plot the base 10 logarithm of the total scattering rate times the solar lifetime, for $\chi \rightarrow \chi^*$ in the left plot and for $\chi^* \rightarrow \chi$ in the right one.

The largest rate for $\chi \rightarrow \chi^*$ scattering is found for the states with low angular momentum and medium energy. These are the particles that have enough energy to travel fairly far out from the solar center. When they fall back into the solar center, they regain a significant amount of kinetic energy, which allows endothermic scattering to take place. Particles with larger energies spend more of their time outside the Sun, which decreases their scattering rate. There are also two regions, one at very low energies and one at very high energies and large angular momenta, where scattering does not take place at all. This can be explained by the fact that the total kinetic energy in collisions taking place in most of the $E$-$L$ plane is supplied almost entirely by the DM particle. The only region where this is not the case is in the very low $E$ region in which DM particle orbits are confined to the solar center. These particles have very low velocities and the energy of nuclei, even though the temperature is high, is not sufficient to provide conditions under which up-scattering can occur. At large $E$ and large $L$, the nuclei are essentially stationary and the DM particles always have low velocities due to their circular orbits, leading to the conclusion that up-scattering is kinematically disallowed also in this region. Even if scattering is allowed, the rates are suppressed due to the DM particles travelling on orbits in which they spend the vast majority of their time outside the Sun.

The scattering of $\chi^* \rightarrow \chi$ is never kinematically suppressed since the process is exothermic. The rates are thus largest for particles that are confined to the solar center, i.e., in the low $E$ and $L$ region. The only suppression in the scattering rate occurs for states at large $E$, which spend more time in less dense regions. The extreme case is thus for very large $E$ and $L$, with highly circular orbits in the outer regions of the Sun, where the density of targets is the lowest, the DM velocity is small, and most of the time is spent outside the Sun. Interestingly, it can also be seen that the rate for exothermic scattering is larger than the rate for endothermic scattering in the entire $E$-$L$ plane. This indicates that the form-factor suppression, which is larger for exothermic scattering, is not as strong as the kinematic suppression of endothermic scattering.

\begin{figure} 
\centering
\includegraphics[width=0.48\textwidth]{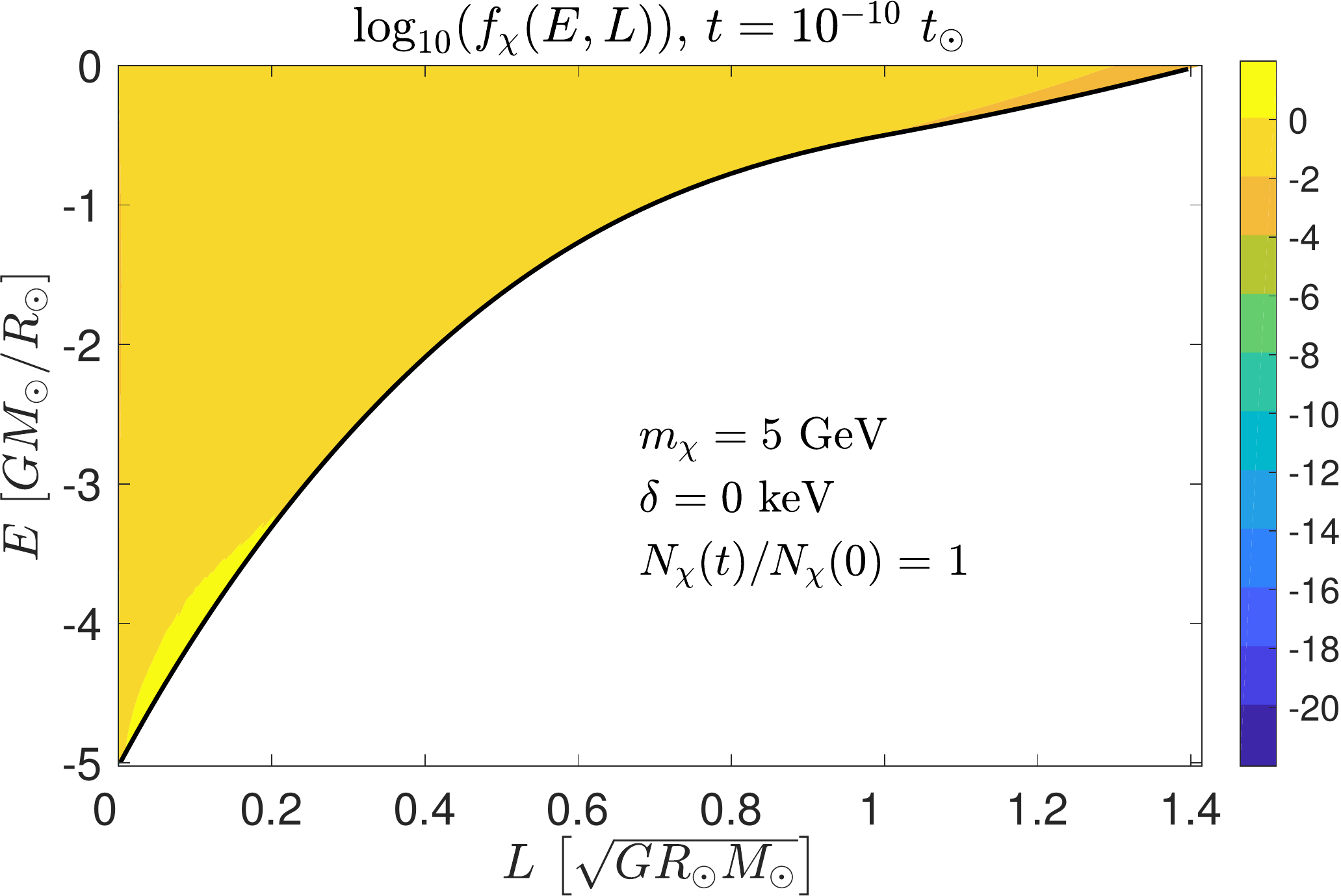}~~
\includegraphics[width=0.48\textwidth]{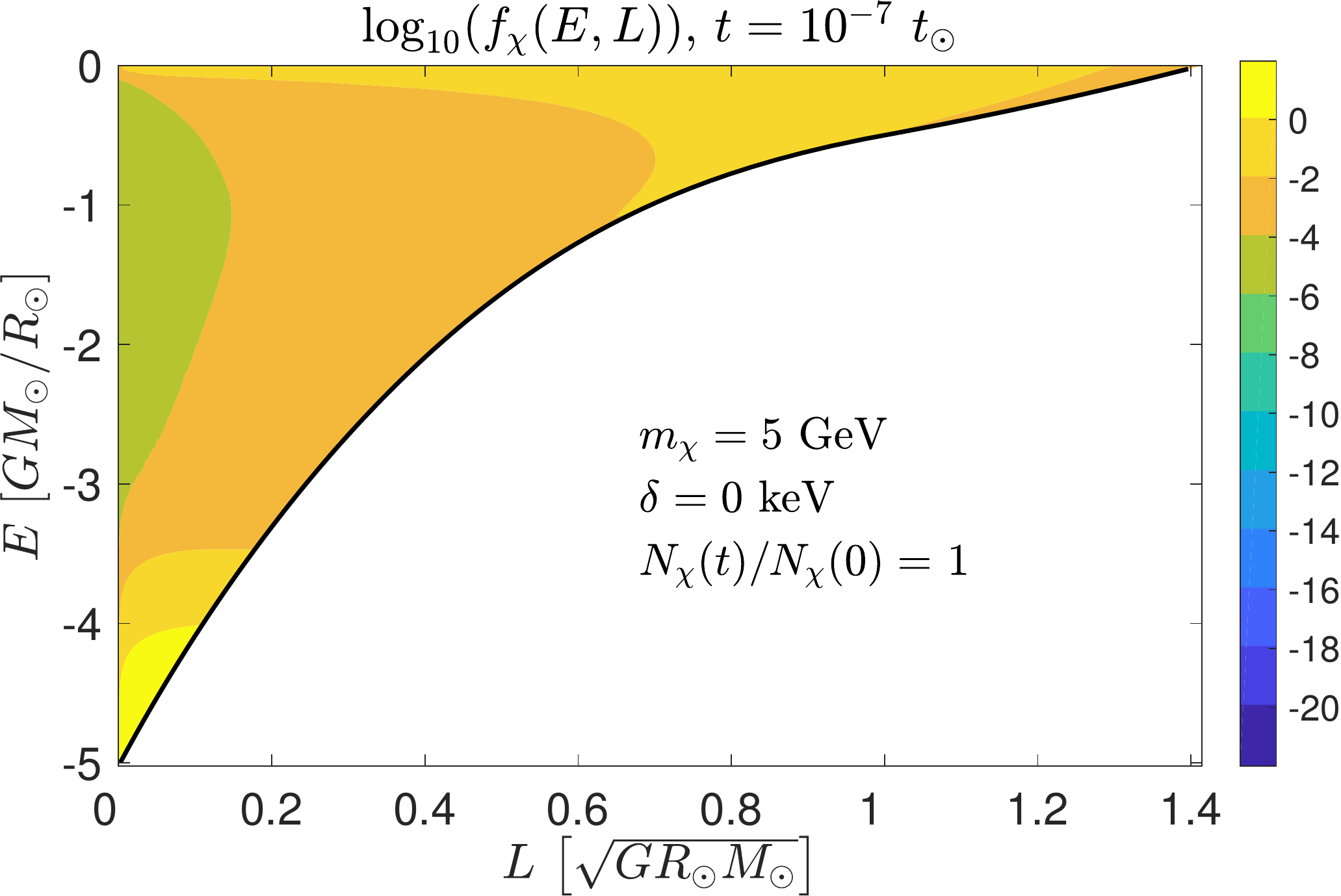}\\
 \vspace{3mm}
\includegraphics[width=0.48\textwidth]{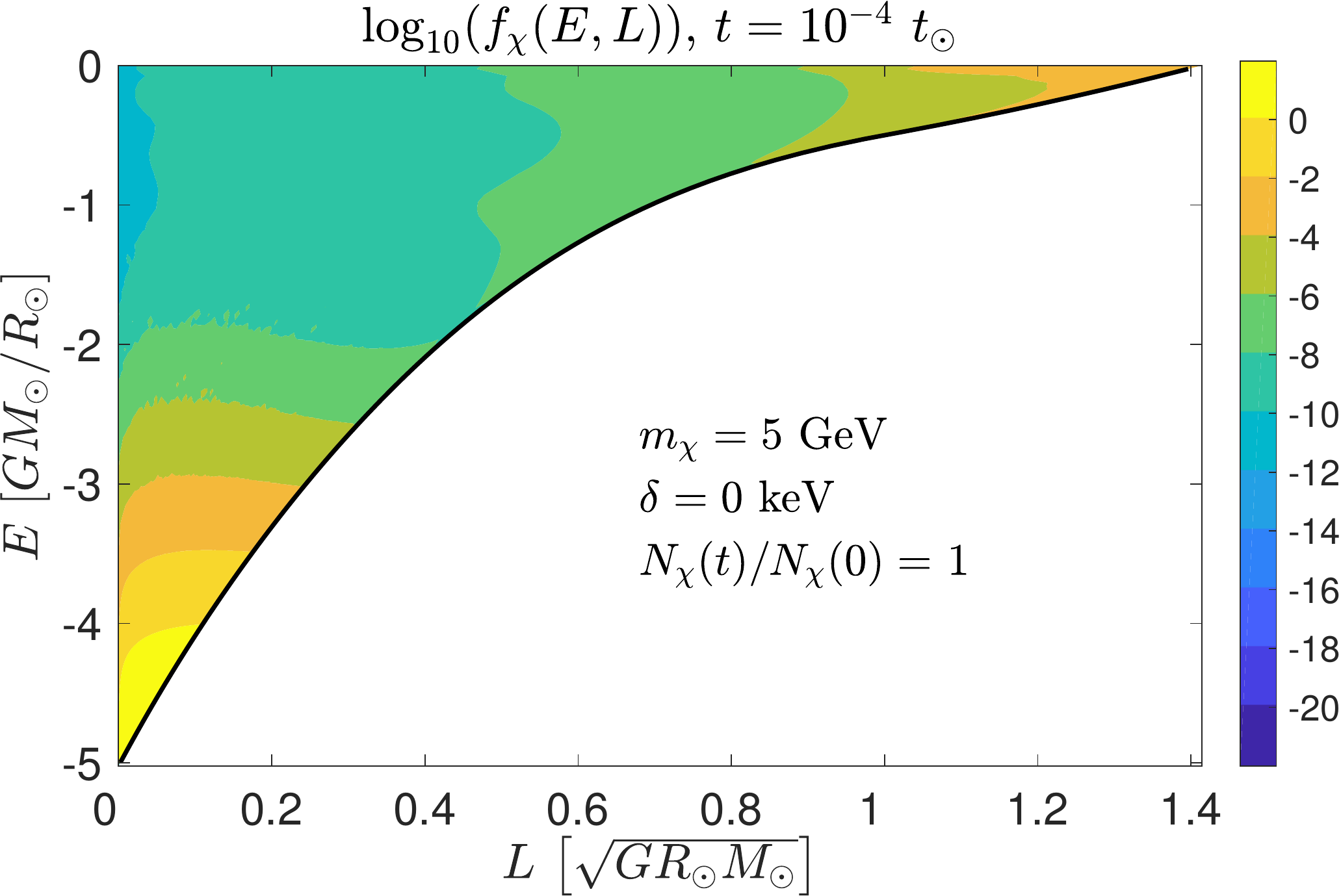}~~
\includegraphics[width=0.48\textwidth]{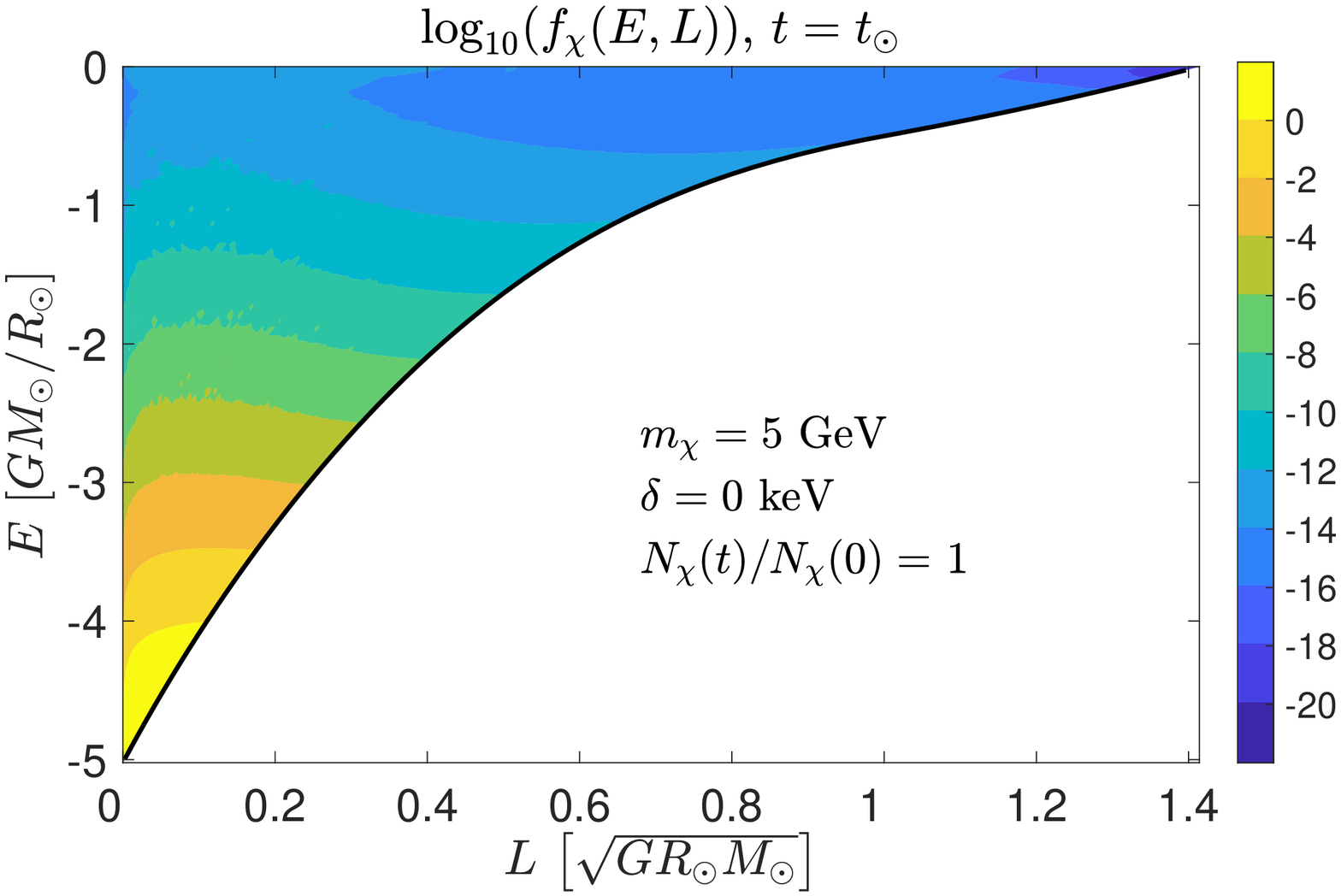} 
\caption{The base 10 logarithm of the distribution of dark matter at various times $t/t_\odot = 10^{-10}, \, 10^{-7}, \, 10^{-4}$ and $1$ for an initial distribution $\vec{f}(0) = \vec{C} \, \Delta t$.  We use elastic scattering with $m_\chi=5$~GeV.} \label{fig:mx5_distevo} 
\end{figure}

Next, we can take a sample of freshly captured DM particles in a time $\Delta t$ which is small enough for no additional scattering to have occurred post capture. The time evolution is then found by solving Eq.~\eqref{eq:governing_eq} neglecting additional capture and annihilation, with the initial distribution $\vec{f}(0) = \vec{C} \Delta t$. The distribution then evolves in time as
\begin{equation} \label{eq:dist_evolve}
\vec{f}(t) = e^{\Sigma t} \, \vec{C} \Delta t \, .
\end{equation}
Figure~\ref{fig:mx5_distevo} shows the base 10 logarithm of the distribution at various times for $m_\chi = 5$~GeV. The distribution accumulates into the lower region of the $E$-$L$ space very rapidly. Taking the scale into account, the distribution comes close to equilibrium at $t \sim 10^{-7} \, t_\odot$, at which point most particles in the Sun have gathered in orbits with very low energies. As time evolves further, there is a constant flow of the few remaining particles at larger $E$ down towards the lower energy orbits. It is also interesting to note that evaporation is negligible over a solar lifetime, i.e., $N_\chi(t_\odot)/N_\chi(0) = 1$.\footnote{The figure at $t = t_\odot$ is in good agreement with Fig.~3.1 in Ref.~\cite{Liang:2016yjf}.}

\begin{figure}
\centering
\includegraphics[width=0.32\textwidth]{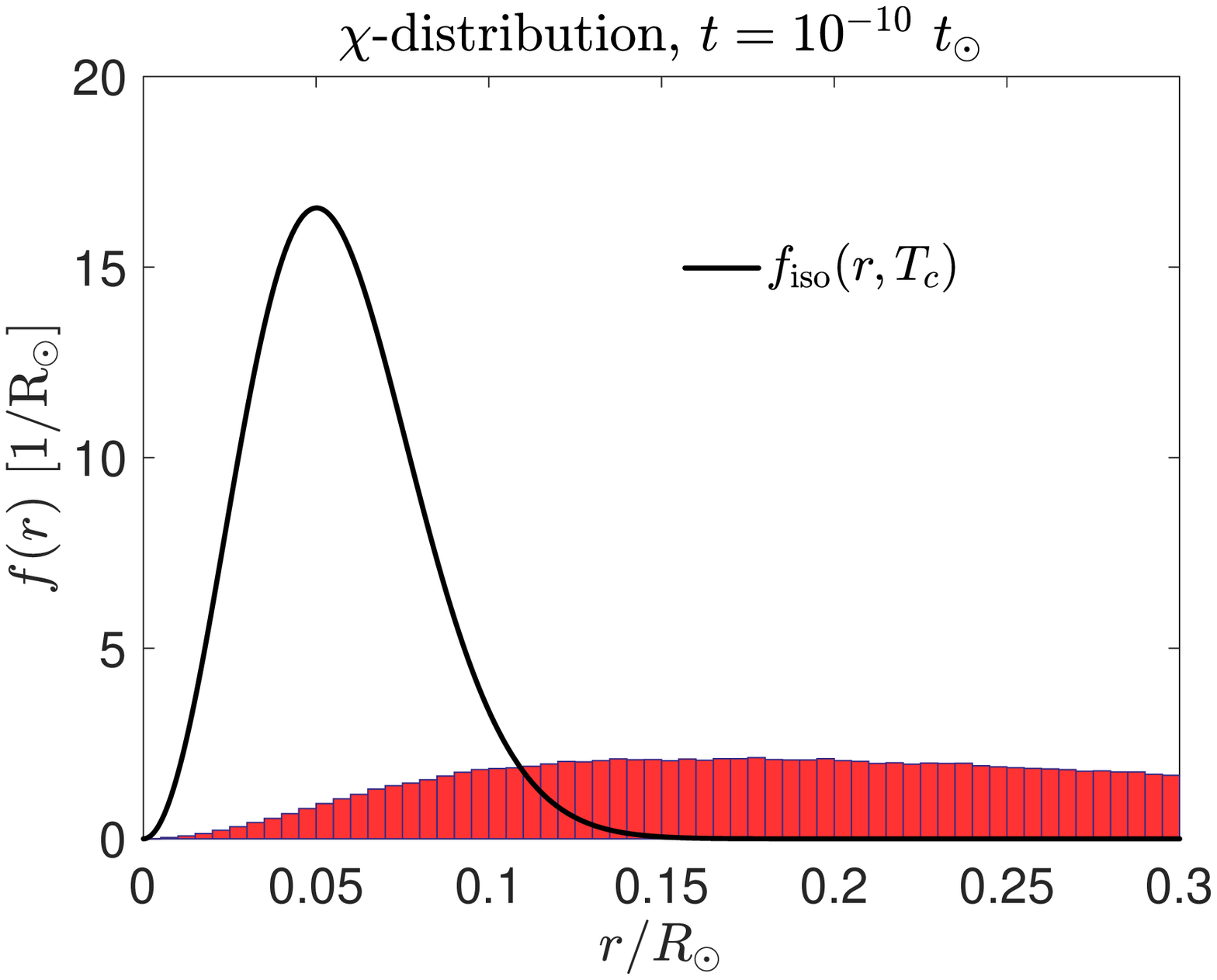}~
\includegraphics[width=0.32\textwidth]{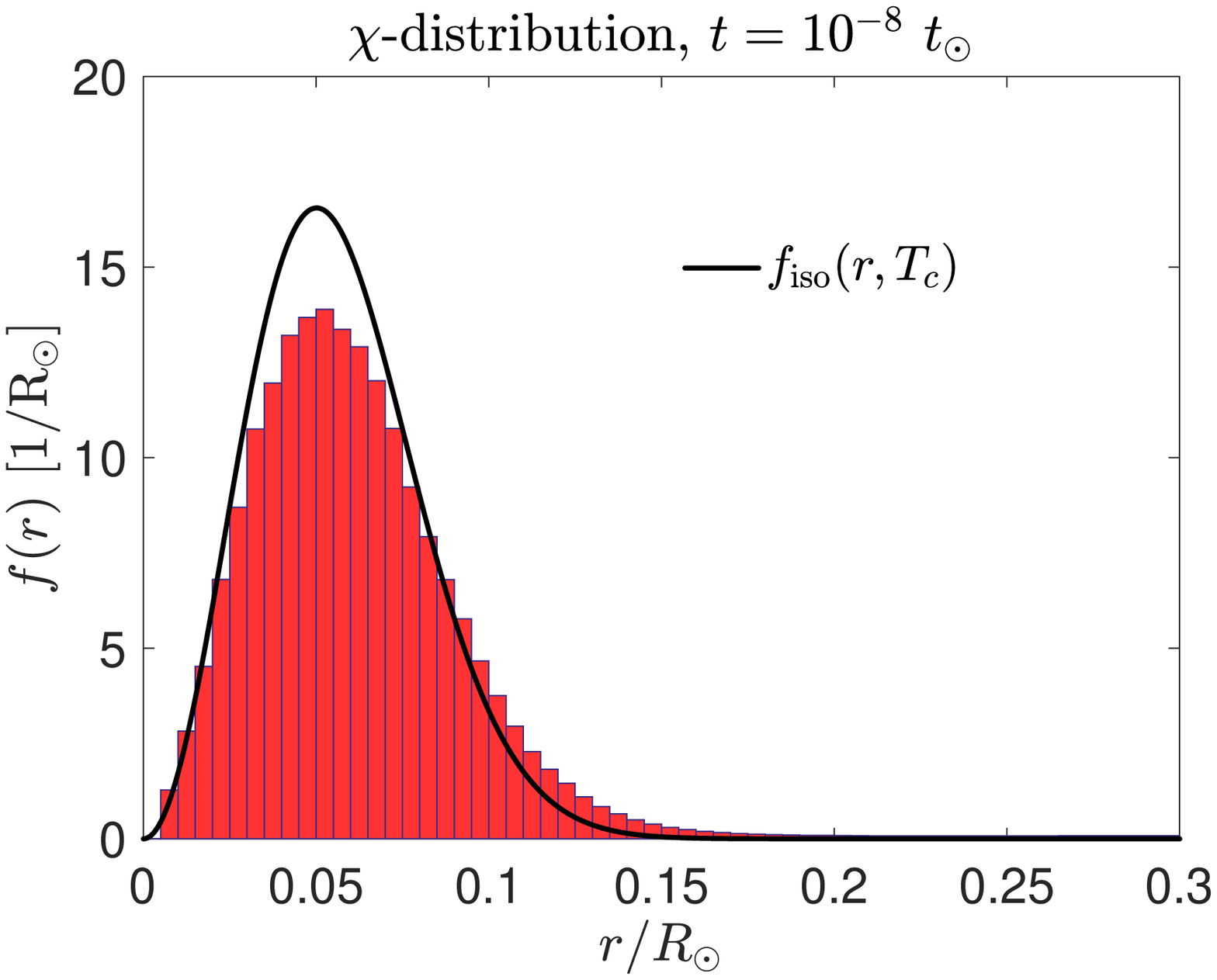}~
\includegraphics[width=0.32\textwidth]{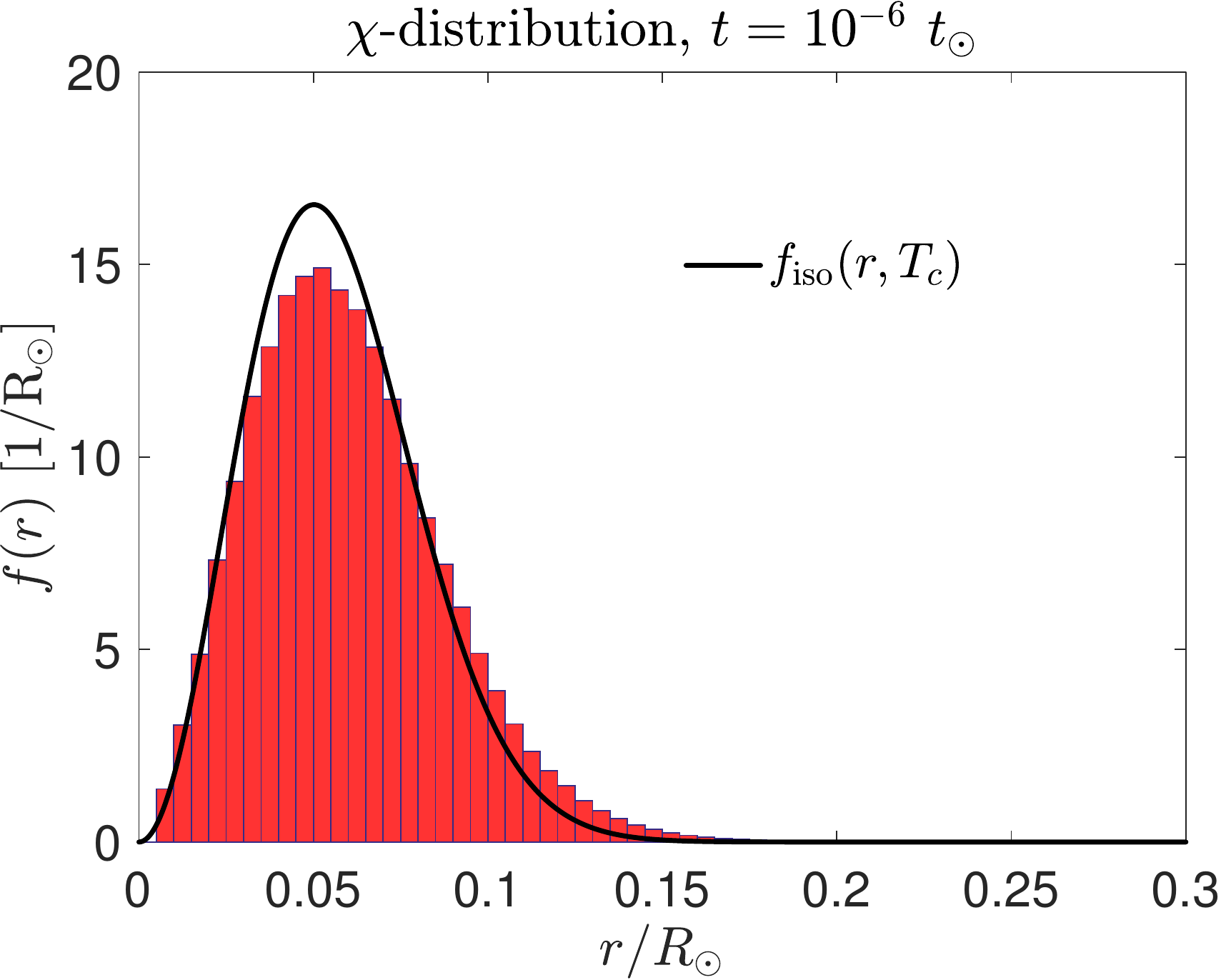}
\caption{The radial distribution of elastic DM at times: $t = 10^{-10} \, t_\odot$ (\emph{left panel}), $t = 10^{-8} \, t_\odot$ (\emph{middle panel}) and $t = 10^{-6} \, t_\odot$ (\emph{right panel}). We fix $m_\chi = 5$~GeV.} \label{fig:mx5_rdists} 
\end{figure}

We now use Eq.~\eqref{eq:r_dist} to translate the distribution in $E$-$L$ space into a radial distribution. The results for elastic scattering are shown in Fig.~\ref{fig:mx5_rdists} for the times $t = 10^{-10} \, t_\odot$ (left), $t = 10^{-8} \, t_\odot$ (middle) and $t = 10^{-6} \, t_\odot$ (right). The distribution is compared to the isothermal one of Eq.~\eqref{eq:iso_dist}, with the angular degrees of freedom integrated over. We see that the distribution has essentially reached equilibrium already at $t = 10^{-8} \, t_\odot$, changing only slightly at $t = 10^{-6} \, t_\odot$. The Boltzmann distribution gives a fairly accurate description of the distribution, although the numerically computed one is slightly shifted towards larger radii, and its peak is not as pronounced.

\begin{figure}
\centering
\includegraphics[width=0.48\textwidth]{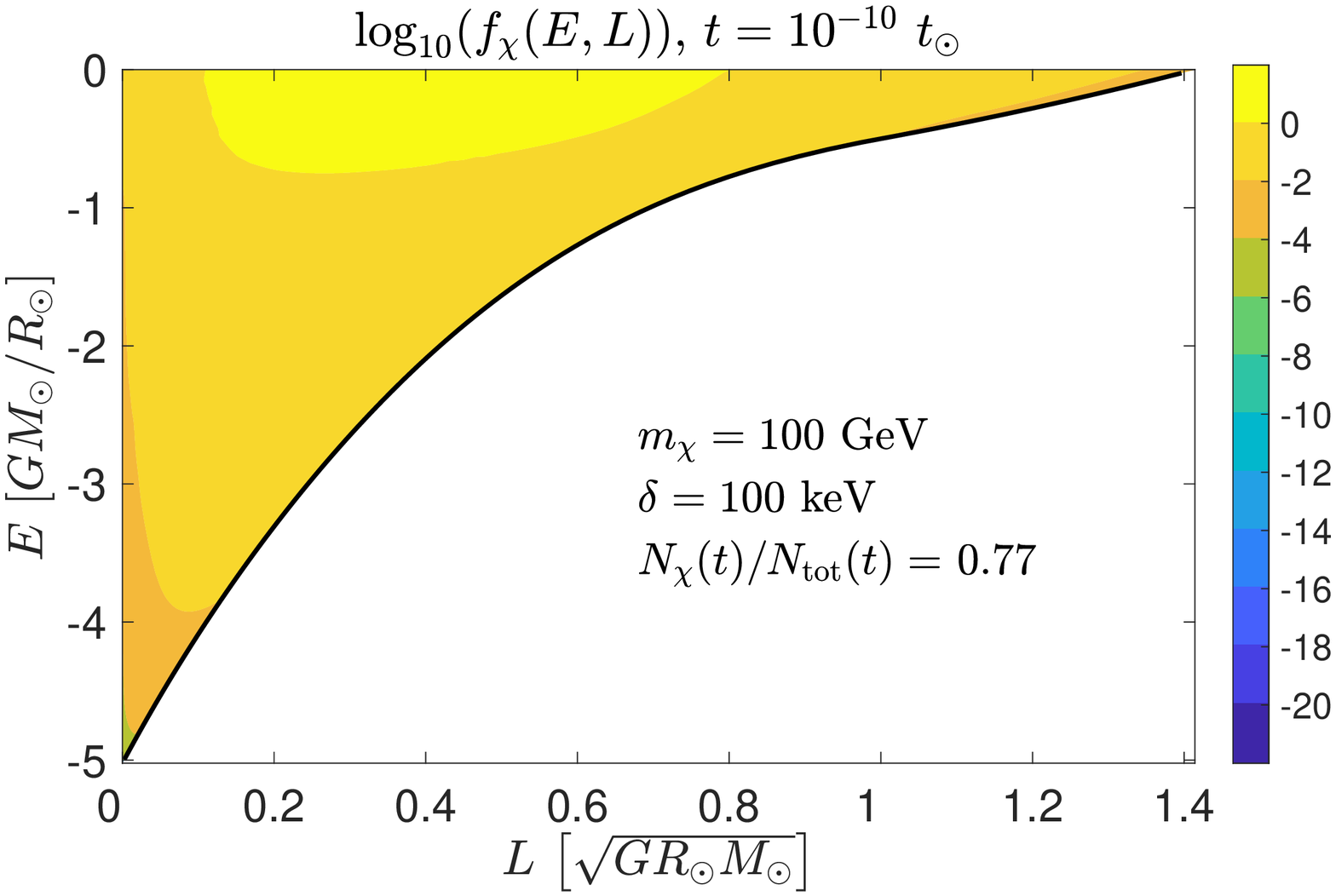}~~
\includegraphics[width=0.48\textwidth]{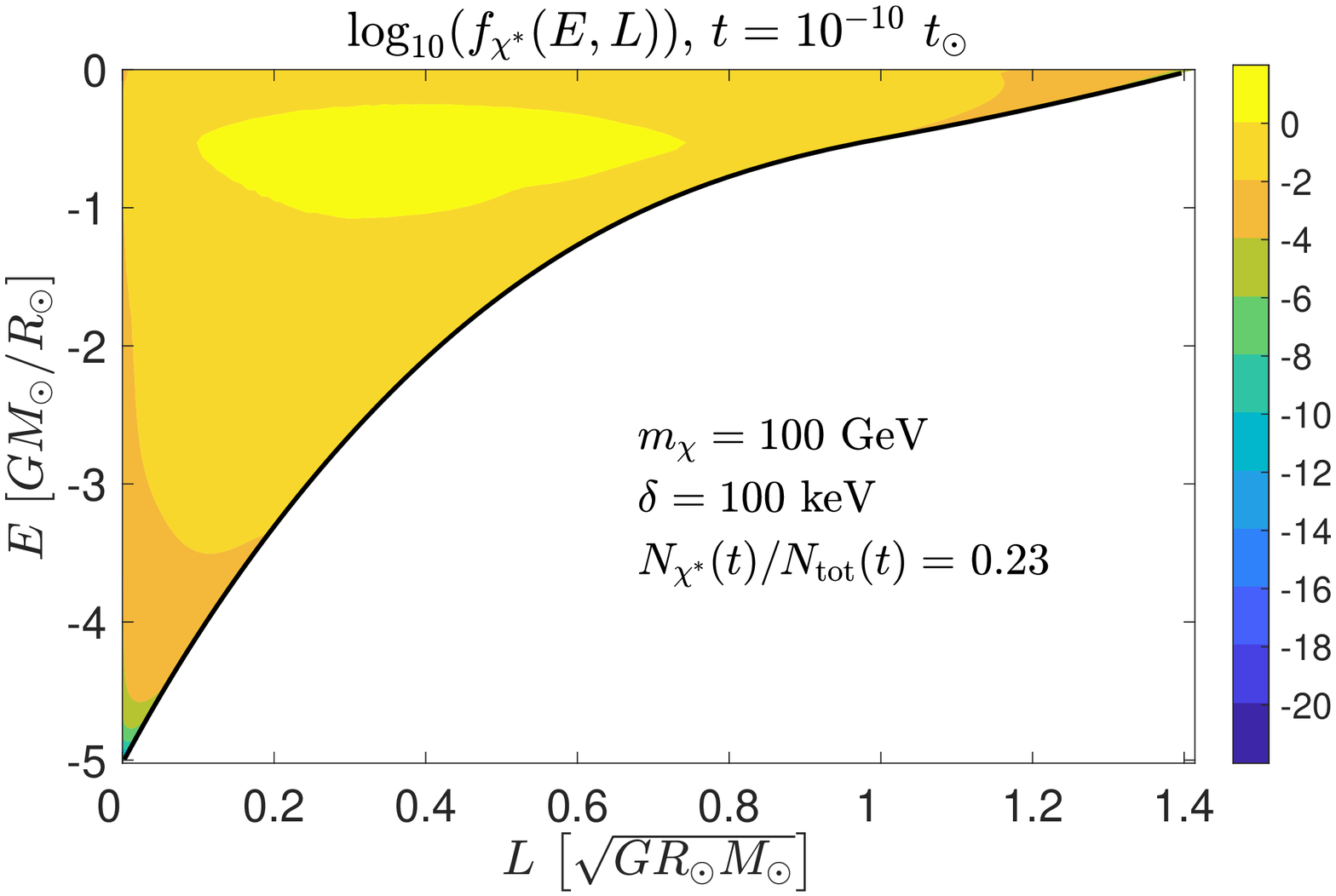}\\
 \vspace{3mm}
\includegraphics[width=0.48\textwidth]{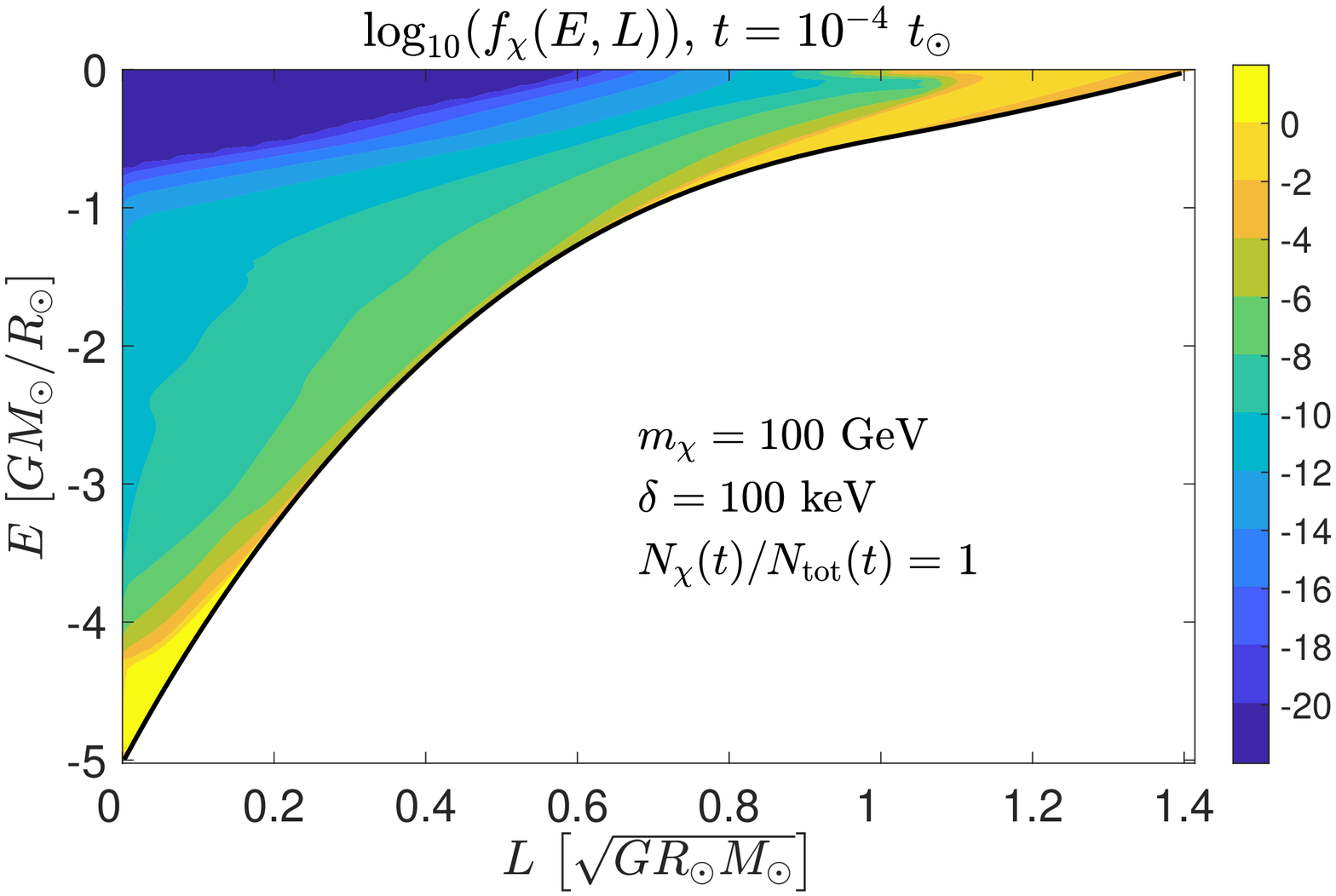}~~
\includegraphics[width=0.48\textwidth]{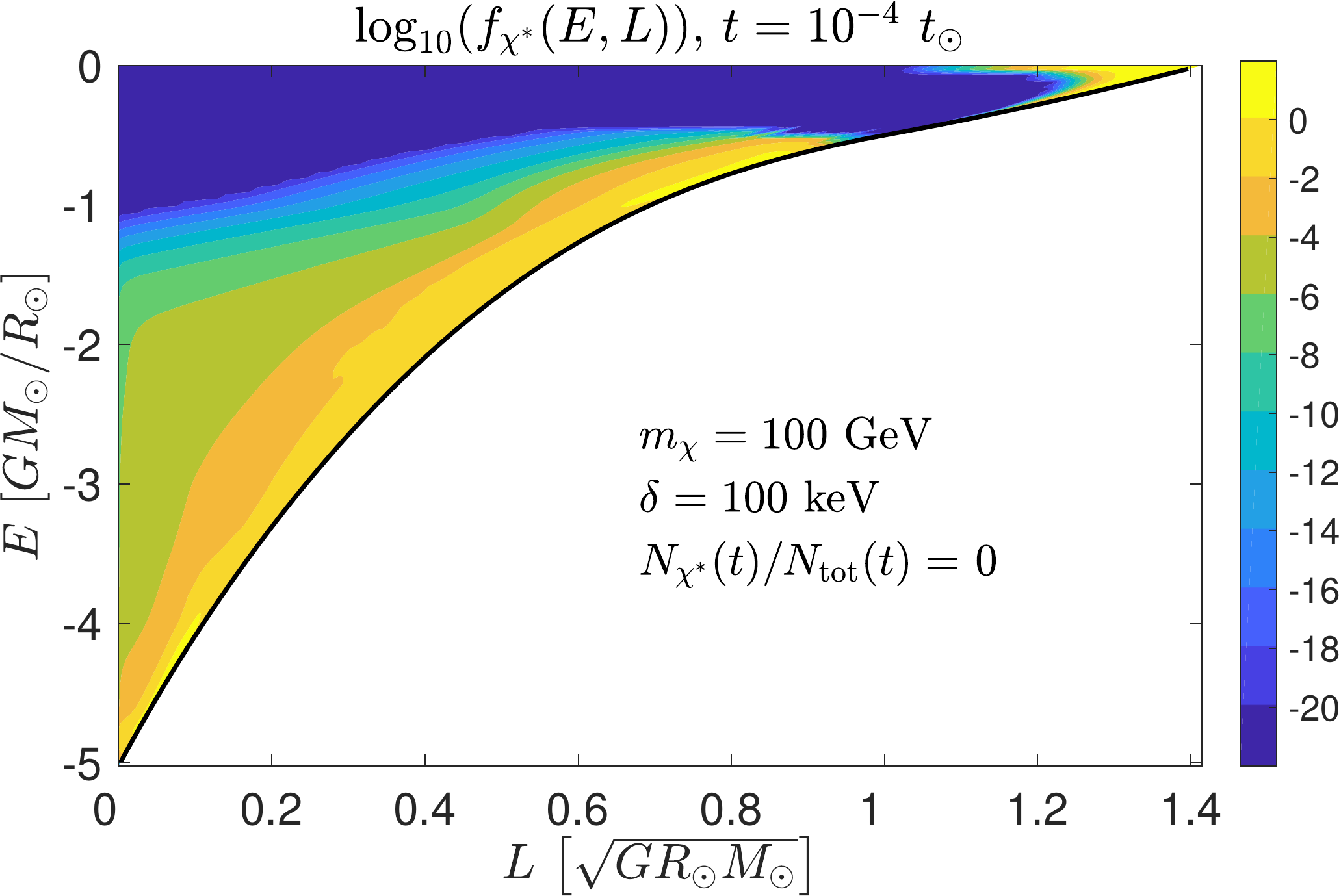}\\
\vspace{3mm}
\includegraphics[width=0.48\textwidth]{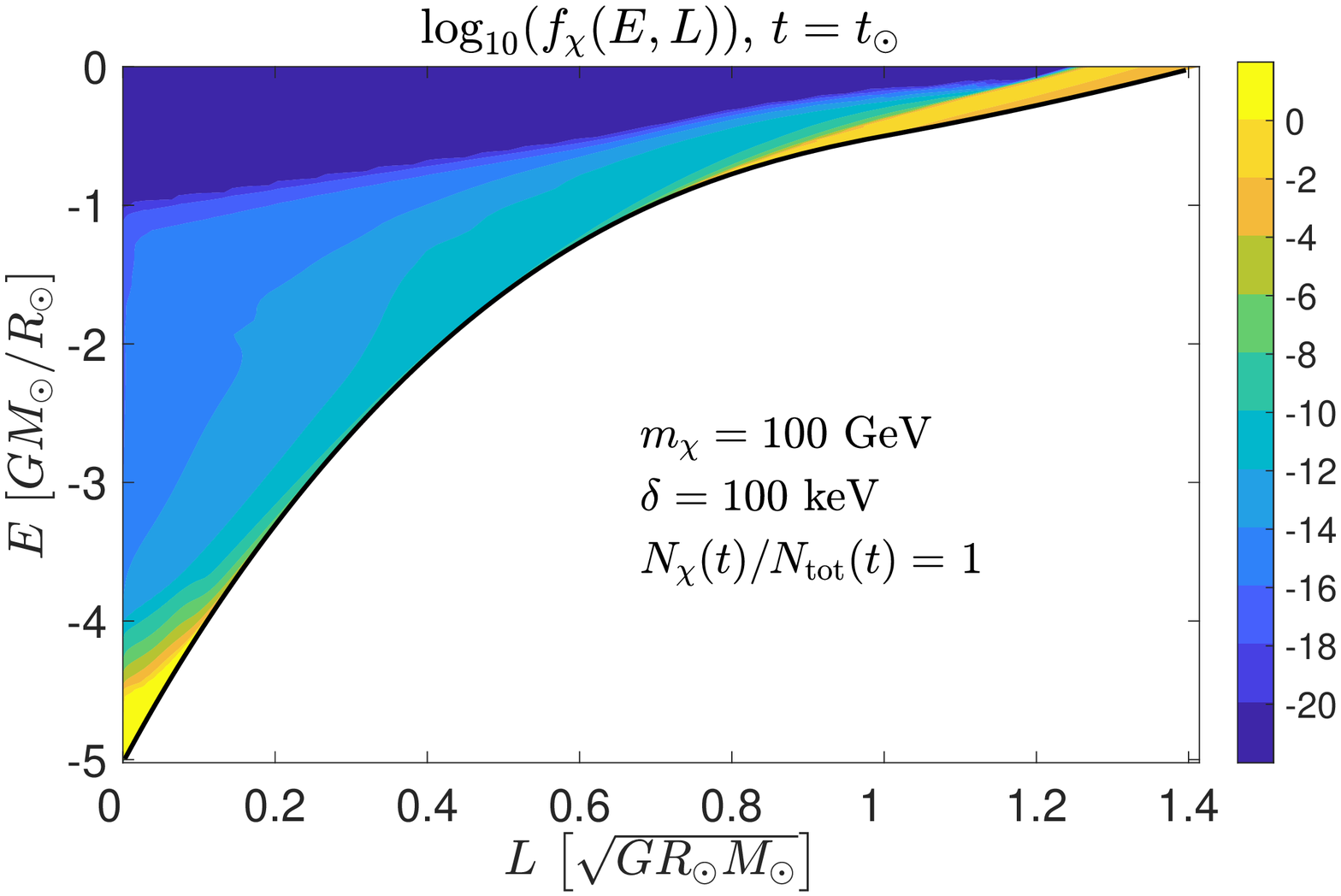}~~
\includegraphics[width=0.48\textwidth]{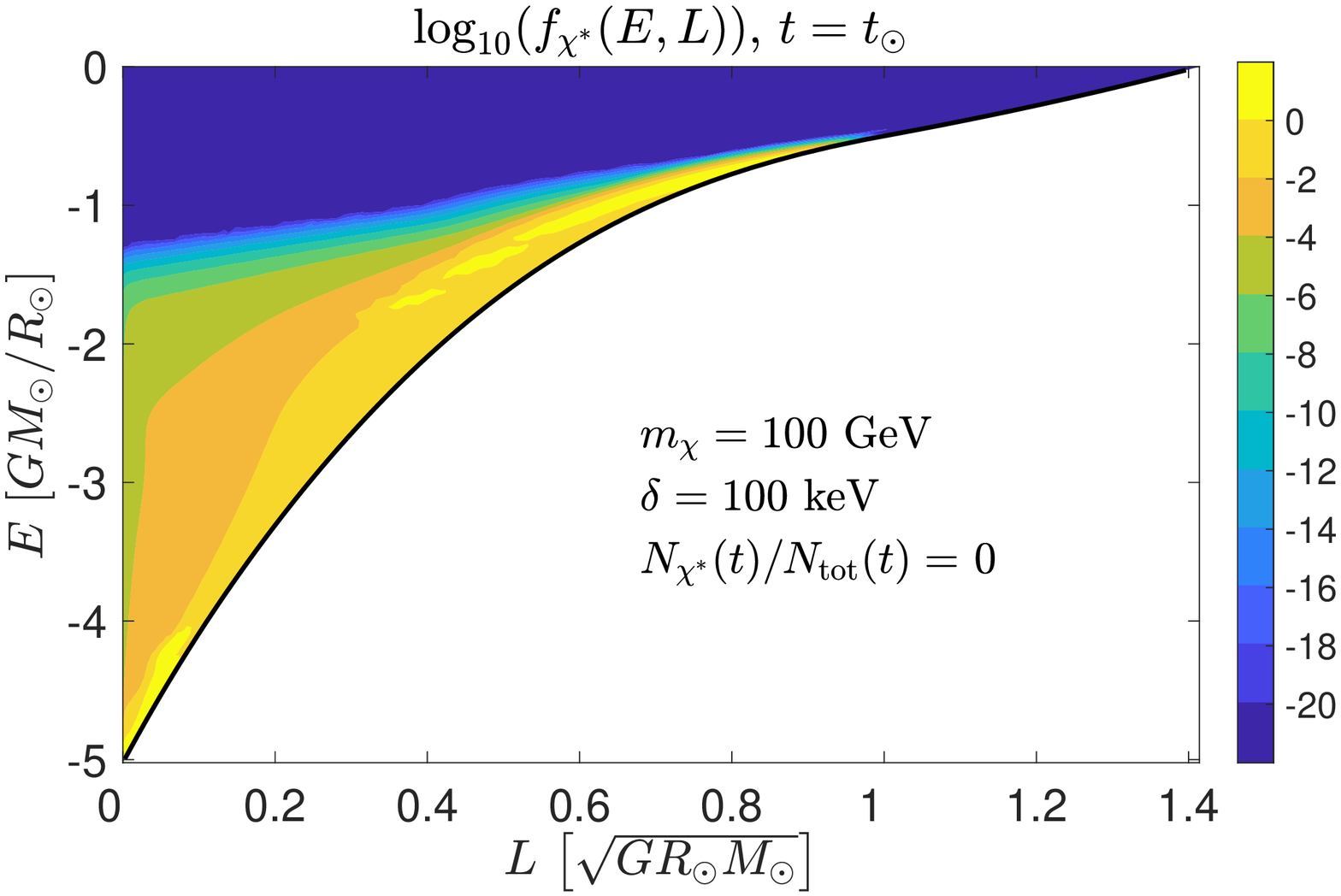}
\caption{The base 10 logarithm of the distribution of DM at various times with an initial distribution given by the capture rate. We use $m_\chi = 100$~GeV and $\delta = 100$~keV. \emph{Left column:} $\chi$ distribution. \emph{Right column:} $\chi^*$ distribution.}  \label{fig:mx100_distevo} 
\end{figure}

Moving on to the case of inelastic DM, we again focus our discussion on the illustrative case of $m_\chi = 100$~GeV and $\delta = 100$~keV. Figure~\ref{fig:mx100_distevo} shows the base 10 logarithm of the $\chi$ ($\chi^*$) distribution in the $E$-$L$ plane in the left (right) plot at various times. The majority of particles in the distribution of $\chi$ have concentrated in the low $E$ region rather quickly as particles at higher $E$ tend to lose energy when scattering and hence fall down the gravitational well. However, there is now also a region at large $E$ and large $L$ that contains a significant number of $\chi$ particles. Rather than particles scattered into this region from other bound orbits, these particles have been primarily captured directly into it, although this is not apparent from Fig.~\ref{fig:mx100_caps} due to the scale used. Even though the capture rate may be low, Fig.~\ref{fig:mx100_scatter} explains the relatively large concentration of particles as up-scattering in this region is kinematically forbidden.

\begin{figure} 
\centering
\includegraphics[width=0.32\textwidth]{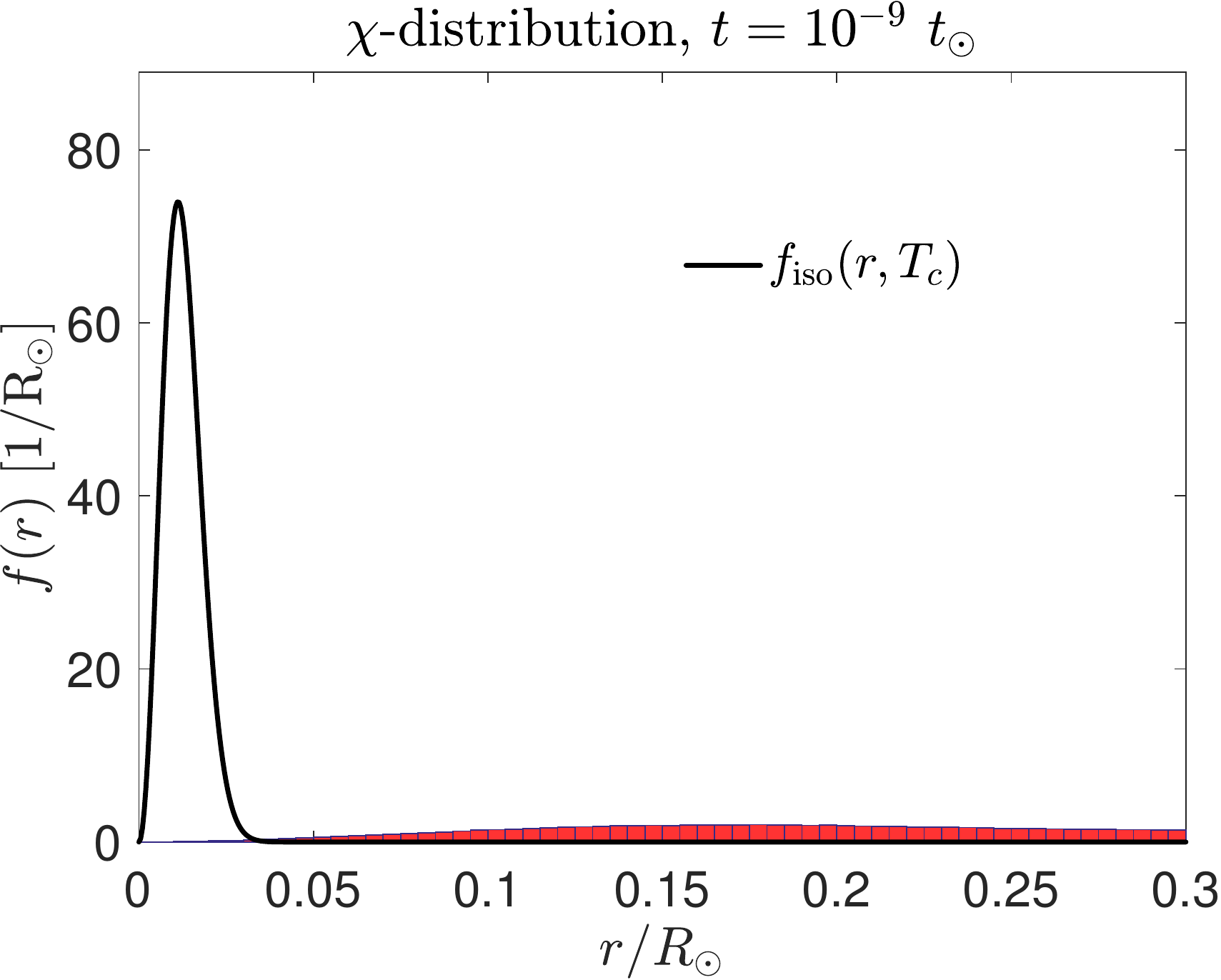}~
\includegraphics[width=0.32\textwidth]{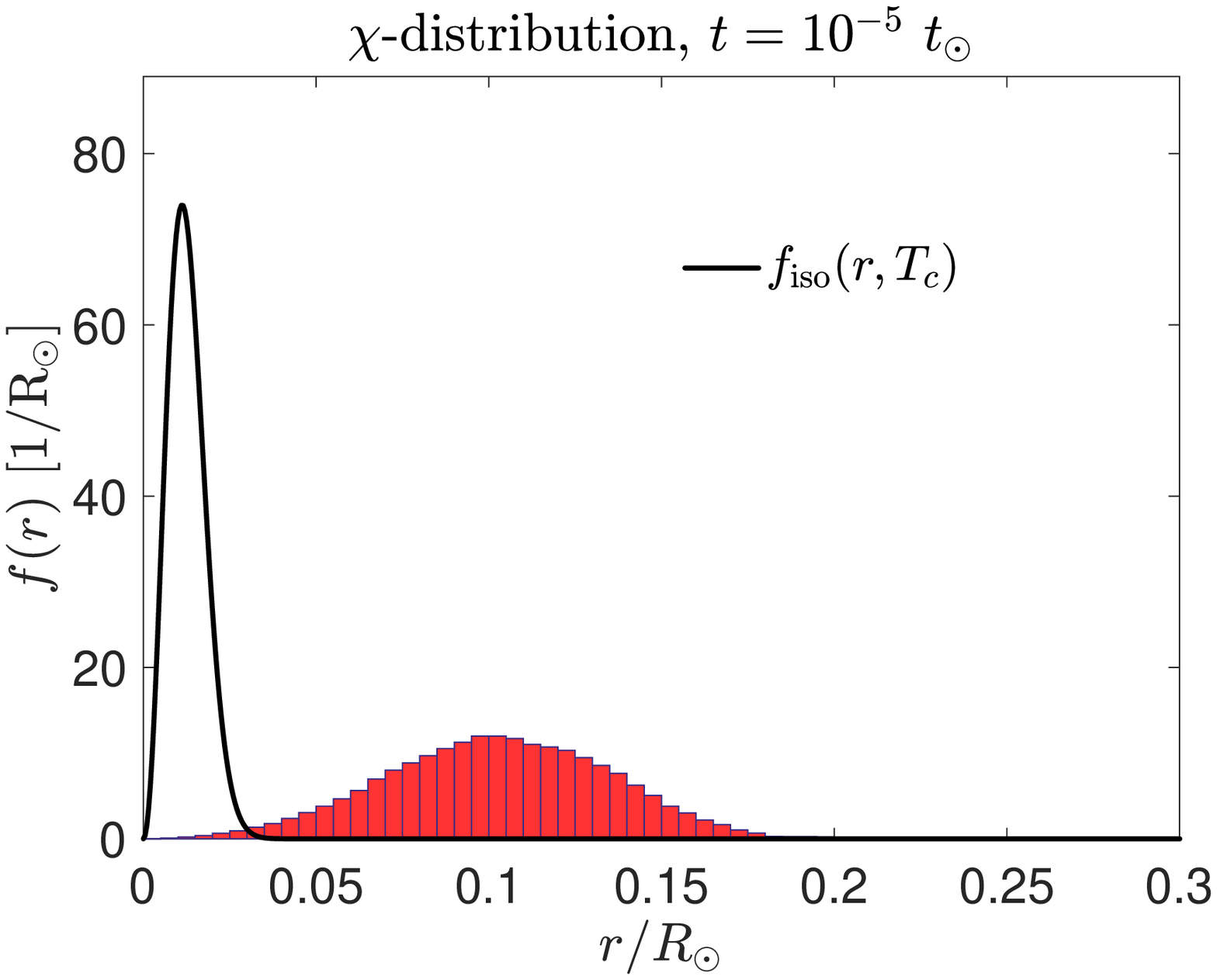}~
\includegraphics[width=0.32\textwidth]{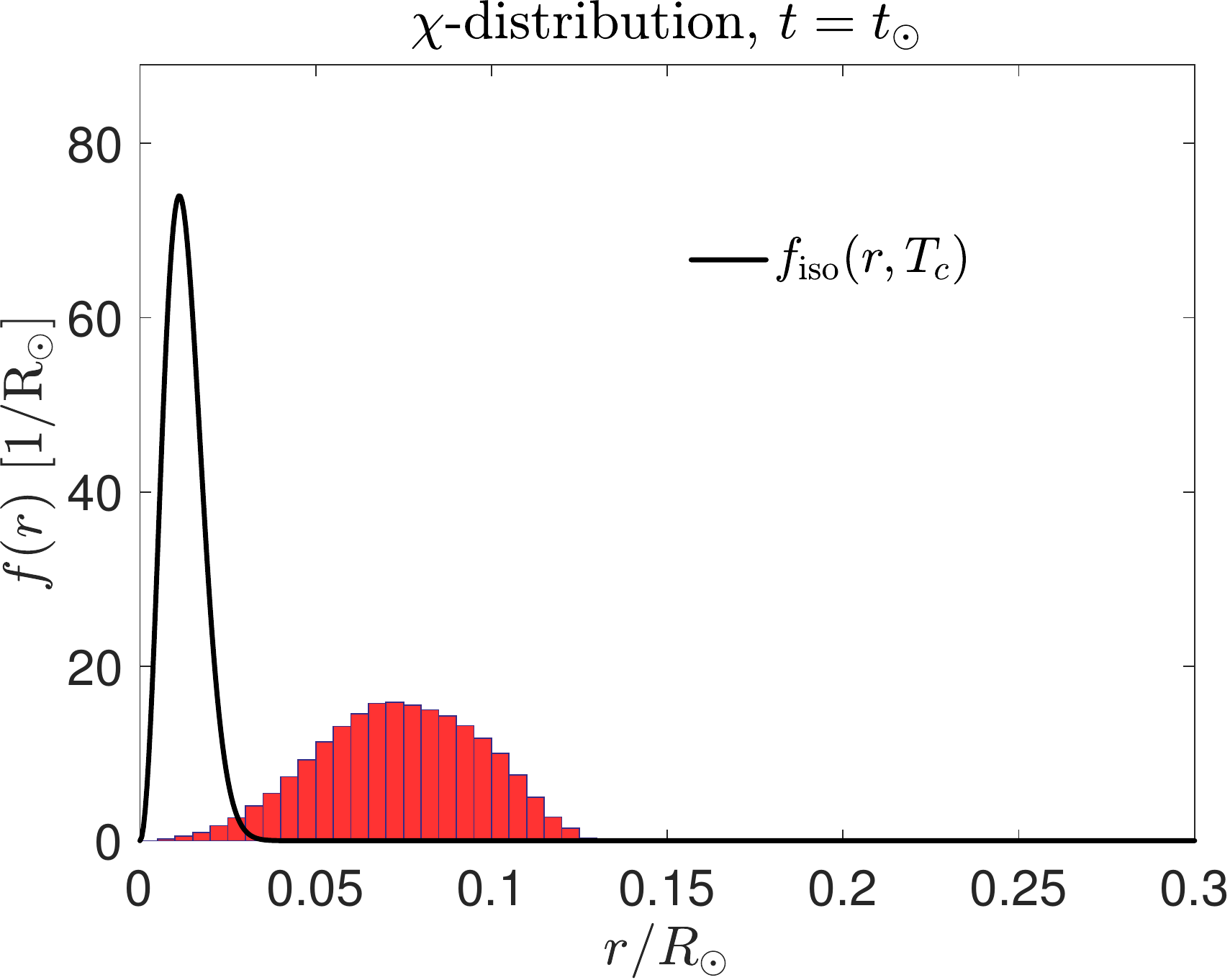}
\caption{The radial distribution of $\chi$ particles, $f_{\rm num}(r)$, at various times $t = 10^{-9} \, t_\odot$ (\emph{left panel}), $t = 10^{-5} \, t_\odot$ (\emph{middle panel}) and $t = t_\odot$ (\emph{right panel}). We use $m_\chi = 100$~GeV and $\delta = 100$~keV.} \label{fig:mx100_rdist}
\end{figure}

We next obtain the radial distributions for inelastic scattering and show the results in Fig.~\ref{fig:mx100_rdist}. At very early times, the distribution extends up to large radii. At $t=10^{-9}\,t_\odot$, a large concentration starts to form, shown below $r/R_\odot \simeq 0.3$. It very slowly moves towards smaller radii, forming a distribution centred at $r/R_\odot \simeq 0.1$ at $t=10^{-5}\,t_\odot$. However, even at $t = t_\odot$ the distribution has yet to reach a stationary state. Another important observation is that the Boltzmann distribution is now a very poor description of the final distribution. This is entirely due to particles being trapped with no possibility of scattering further, in particular those in the region with low $E$, which are the ones that contribute to $f_{\rm num}(r)$ at smaller radii. Due to the circular orbits of particles at large $E$ and $L$, their contribution to the radial distribution is at significantly larger radii ($r \gtrsim 0.6$~$R_\odot$) than that shown in Fig.~\ref{fig:mx100_rdist}. This contribution is not significant due to their low abundance relative to the distribution close to the solar center. Since the number of $\chi^*$ particles is completely negligible, the total DM distribution is practically identical to the $\chi$ distribution.

\begin{figure} 
\centering
\includegraphics[width=0.48\textwidth]{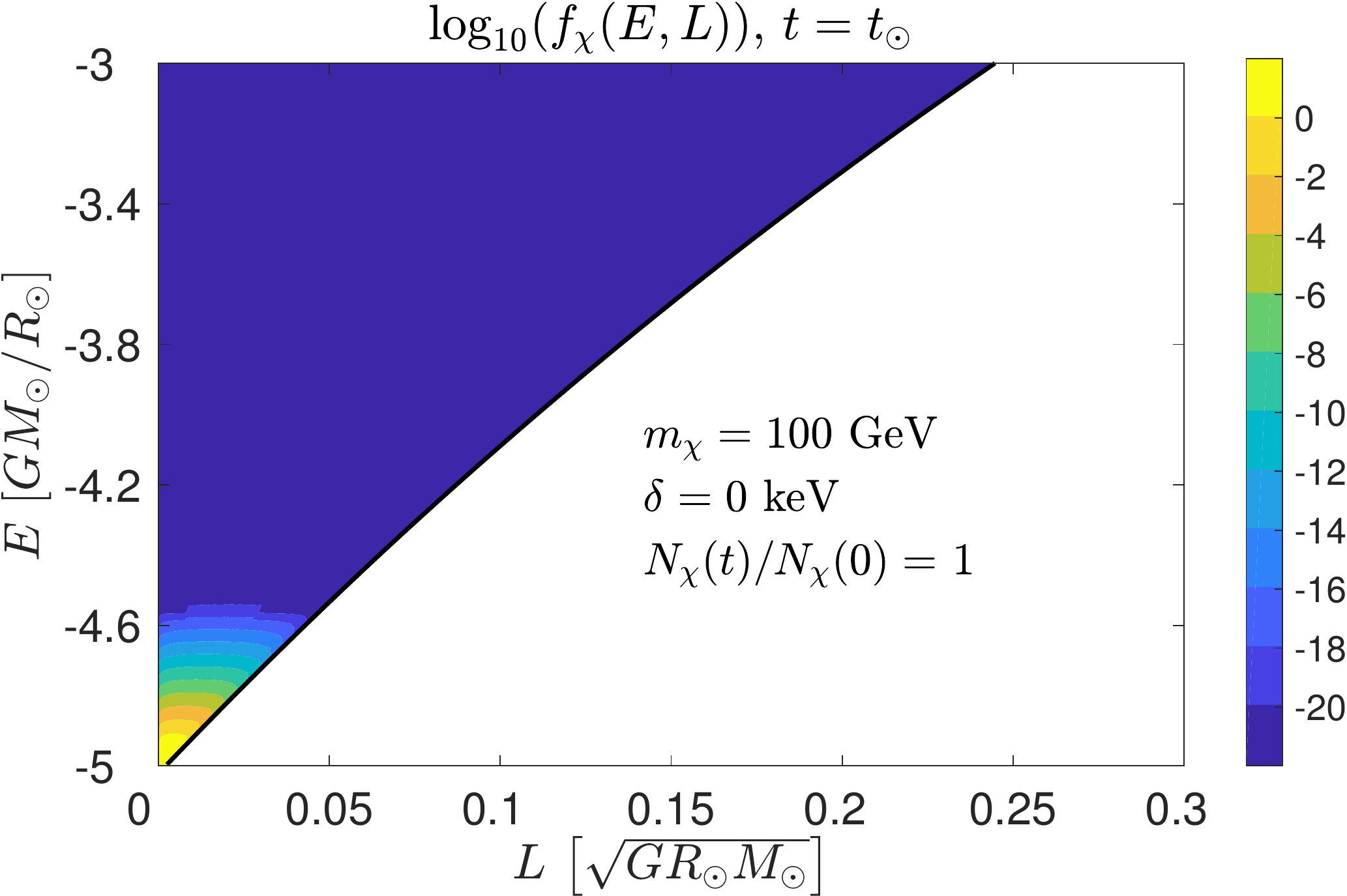}~~
\includegraphics[width=0.48\textwidth]{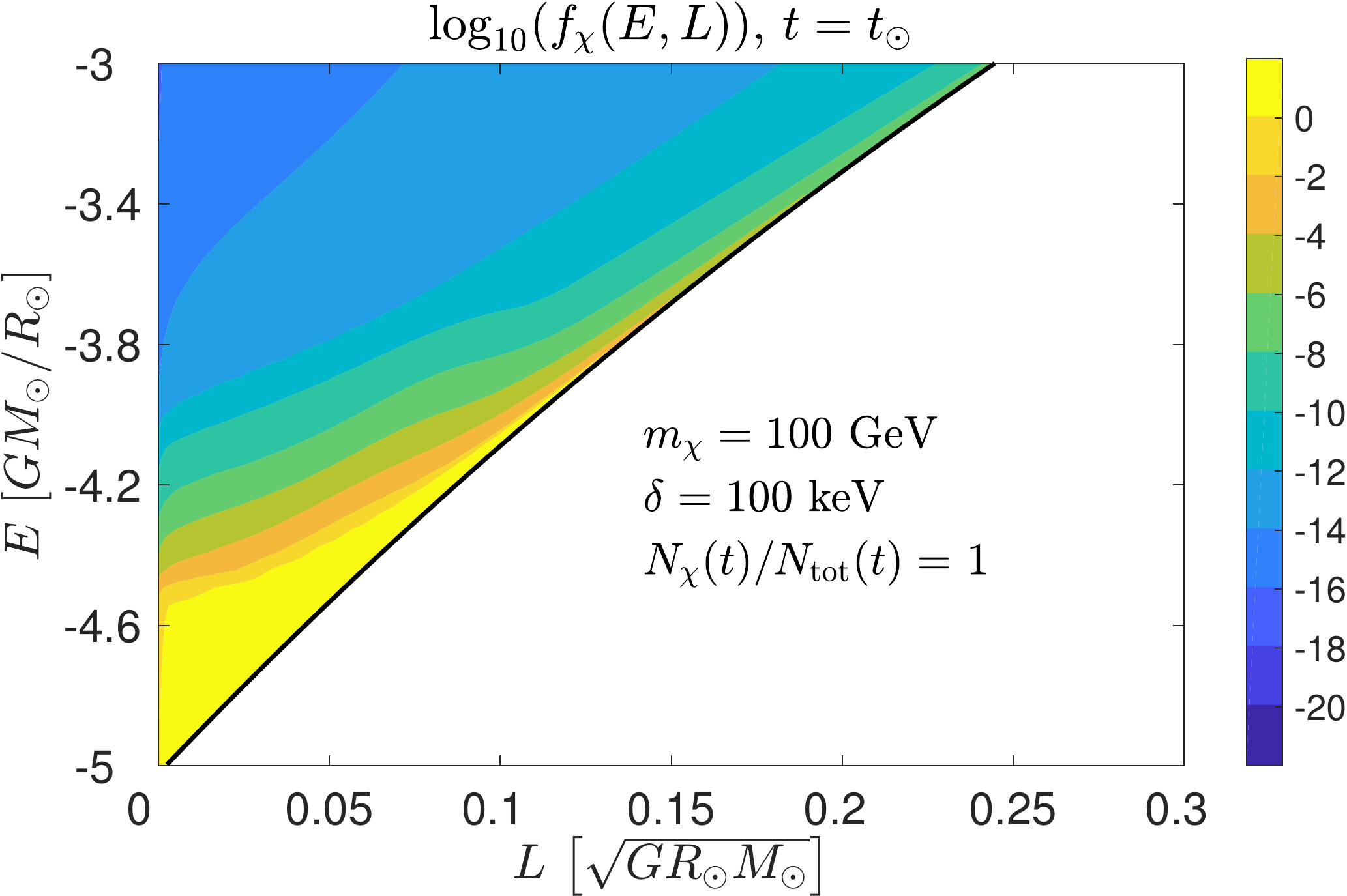}
\caption{The base 10 logarithm of the dark matter distribution at $t=t_\odot$, with an initial distribution given by the capture rate. We use $m_\chi = 100$~GeV. \emph{Left panel:} Elastic scattering. \emph{Right panel:} Inelastic scattering with $\delta = 100$~keV.} \label{fig:mx100_grid_elast}
\end{figure}

Finally, it is interesting to compare the DM distributions at $t = t_\odot$ between the elastic and the inelastic cases. In Fig.~\ref{fig:mx100_grid_elast} we show, for $m_\chi = 100$~GeV, the elastic case (left panel) and the inelastic one with $\delta = 100$~keV (right panel). As can be seen, the distribution is (as expected) extremely concentrated towards the central region of the Sun. One can also observe that no evaporation has taken place. The distribution is in fact concentrated into so few states that a reliable radial distribution cannot be derived unless a significant increase in the number of low $E$ states used in the simulation is made. This is a problem that also appears for inelastic DM when $\delta$ is small enough, which prevents us from calculating the annihilation rate for arbitrary low values of $\delta$ using the method described here.

\subsection{Annihilation and evaporation}

In order to investigate the effects that the altered distribution has on the annihilation rate, we use Eq.~\eqref{eq:dist_greens} to calculate the number density functions at $t = t_\odot$. This distribution contains information on the total number of particles in the Sun and therefore provides a more realistic distribution. In fact, sets of particles that are captured at different times are distributed differently in the Sun at $t = t_\odot$ and thus contribute differently to the overall distribution. Of course, the number of particles that have evaporated is also affected by the amount of time passed since they were originally captured.

\begin{figure}
\centering
\includegraphics[width=0.48\textwidth]{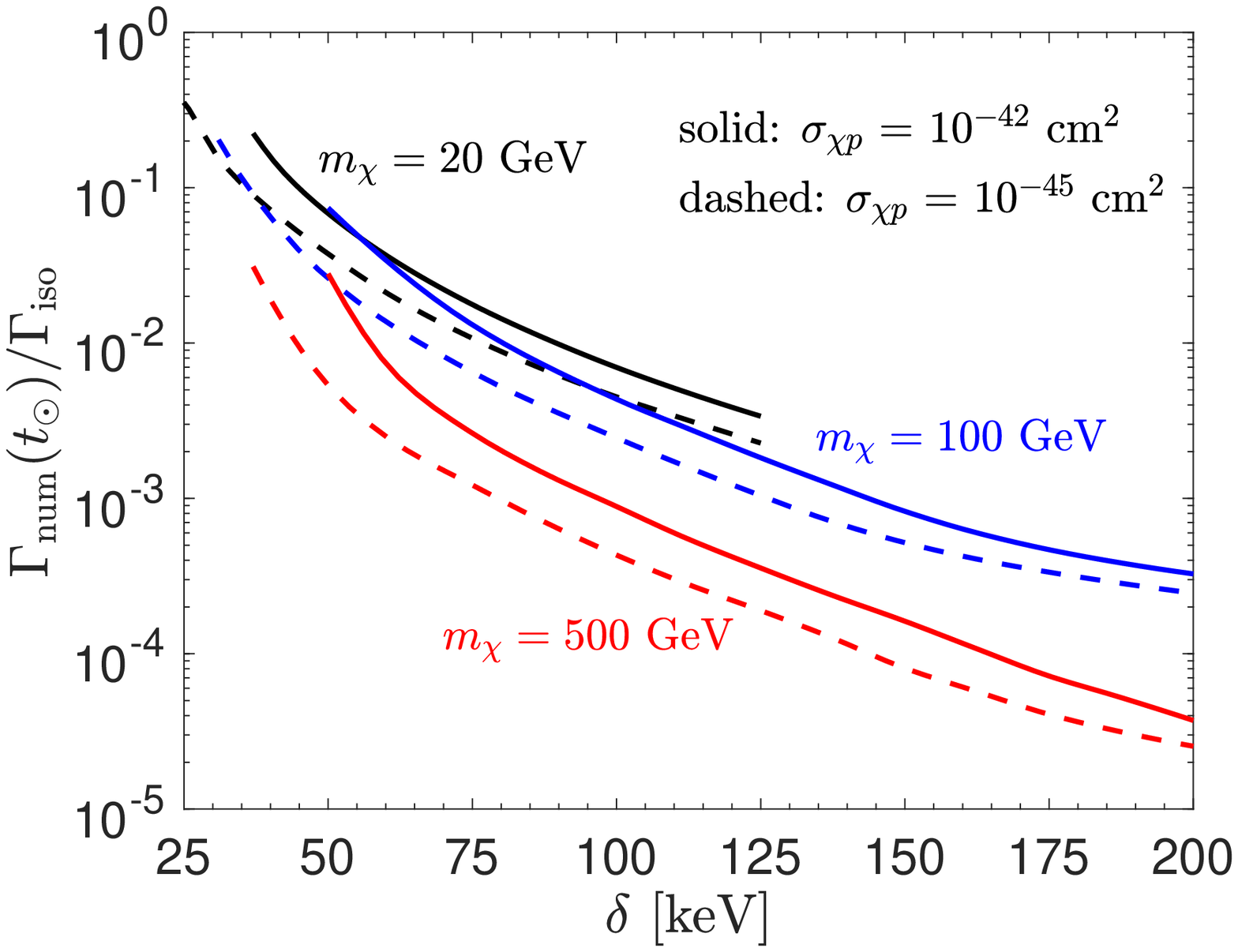}~~
\includegraphics[width=0.48\textwidth]{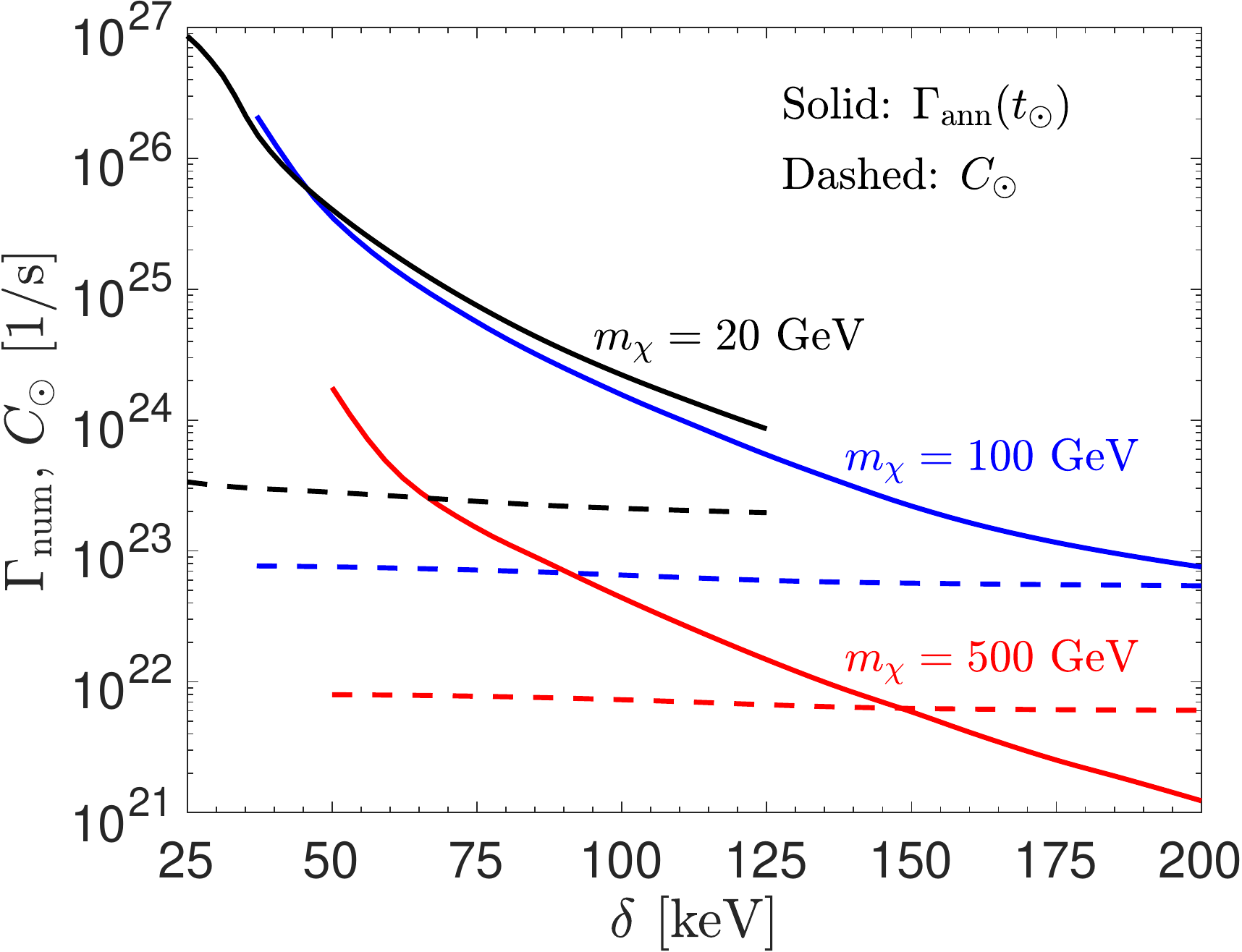}
\caption{\label{fig:anni} \emph{Left panel:} Ratio of the annihilation rates calculated with the numerically obtained distribution and with the isothermal distribution using $\sigma_{\chi p} = 10^{-42}$~cm$^2$ (solid lines) and $\sigma_{\chi p} = 10^{-45}$~cm$^2$ (dashed lines). \emph{Right panel:} Annihilation rates (solid lines) and capture rates (dashed lines) for the numerically computed distributions using $\sigma_{\chi p} = 10^{-42}$~cm$^2$ and $\langle \sigma_{\rm ann} v \rangle = 3 \cdot 10^{-26}$~cm$^3$/s.
In both panels we show results for $m_\chi = 20$~GeV (black), $m_\chi = 100$~GeV (blue) and $m_\chi = 500$~GeV (red).}
\end{figure}

We now calculate distributions of inelastic DM at $t = t_\odot$ for different masses and cross sections. In the left panel Fig.~\ref{fig:anni} we show the ratio of the numerically calculated annihilation rate to the isothermal one, computed using Eq.~\eqref{eq:anni_comp}, as a function of $\delta$. Again, the problem of deriving radial distribution functions for low values of $\delta$ due to the $E$-$L$ discretisation used is encountered, which is why in Fig.~\ref{fig:anni} we only show annihilation rates for larger values of $\delta$, where the problem is avoided. The results are shown for two scattering cross sections: $\sigma_{\chi p} = 10^{-42}$~cm$^2$ (solid lines) and $\sigma_{\chi p} = 10^{-45}$~cm$^2$ (dashed lines), and for three different DM masses: $m_\chi=20$~GeV (black lines), $m_\chi = 100$~GeV (blue lines), and $m_\chi = 500$~GeV (red lines). The annihilation rate is severely suppressed for large values of $\delta$. The reason is that, as $\delta$ increases, so do the regions in which additional scattering of $\chi$ is kinetically forbidden, which in turn leads to a more diluted DM distribution. 

One can also observe that the suppression for $\sigma_{\chi p} = 10^{-42}$~cm$^2$ is not as severe as the one for $\sigma_{\chi p} = 10^{-45}$~cm$^2$. This implies that the distribution does not reach a steady state within a solar lifetime, as considering different scattering cross sections is equivalent to considering different times. Note that the number of captured particles depends on the scattering cross section. However, $N$ drops out from the ratio $\Gamma_{\rm num} (t_\odot)/\Gamma_{\rm iso}$ and therefore the larger ratio for the larger scattering cross section is not due to the total number of DM particles, but only to their different distributions.

Next, we use the number density distributions, computed with the same parameters as before, to calculate the actual annihilation rate using Eq.~\eqref{eq:anni_calc}. We assume s-wave annihilation and assign the thermal averaged cross section the value $\langle \sigma_{\rm ann} v \rangle = 3 \cdot 10^{-26}$~cm$^3$/s. The results are shown in the right panel of Fig.~\ref{fig:anni}, where the annihilation rates (solid lines) are compared to the solar capture rates (dashed lines). We only show the results for $\sigma_{\chi p} = 10^{-42}$~cm$^2$, keeping in mind that the smaller the scattering cross section the lower the annihilation rate. In order to understand the comparison between $C_\odot$ and $\Gamma_{\rm num}$ one should now recall two assumptions that have been made. First, we assumed that the distribution is spherically symmetric, so that the overall annihilation rate would decrease if some orbital plane was preferred. Second, we have assumed that no annihilation has taken place, which means that the actual annihilation rate is overestimated. Thus $\Gamma_{\rm num}$ should be regarded as an upper bound on the annihilation rate under the assumption of a spherically symmetric distribution. 

As can be seen in the right panel of Fig.~\ref{fig:anni}, the upper limit on the annihilation rate generally exceeds the capture rate in most of the considered parameter space. The only case where equilibrium between capture and annihilation has definitely not taken place is for $m_\chi = 500$~GeV and $\delta \gtrsim 150$~keV. The upper bound on the annihilation rate exceeding the solar capture rate does not imply that equilibrium between capture and annihilation has occurred. However, annihilation is very inefficient until a large enough abundance of DM has been accumulated. When the upper limit on the annihilation rate is much greater than the capture rate, equilibrium between the two can be assumed.

Finally, we are interested in knowing how much evaporation affects the total number of DM particles. In the elastic case, it is generally accepted that DM particles with masses below $m_\chi \sim 3-4$~GeV evaporate before being able to annihilate after they are captured~\cite{Gould:1987ju,Busoni:2013kaa}. The situation is not at all as clear in the case of inelastic DM, due to the absorption and release of energy as the DM particle scatters back and forth between the heavier and lighter states. Evaporation is thus a cause for concern, in particular for sizeable $\delta$.

\begin{figure}
\centering
\includegraphics[width=0.5\textwidth]{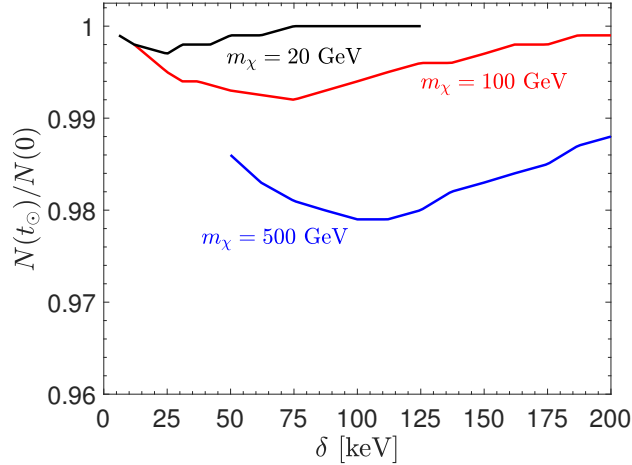}
\caption{\label{fig:evap} Ratio between the total number of particles, after the distribution has evolved for a solar lifetime, and the initial number of particles. Results for different dark matter masses are shown: $m_\chi = 20$~GeV (black), $m_\chi = 100$~GeV (red) and $m_\chi = 500$~GeV (blue).}
\end{figure}

In Fig.~\ref{fig:evap} we show the total number of particles $N(t_\odot)$ in the Sun at $t = t_\odot$ divided by its initial value $N(0)$, where $N(t)$ is calculated using Eq.~\eqref{eq:dist_evolve} for different DM mass values: $m_\chi = 20$~GeV (black line), $m_\chi = 100$~GeV (red line) and $m_\chi = 500$~GeV (blue line). We see that there is a value of the splitting $\delta_{\rm max} (m_\chi)$ where the evaporation rate reaches a maximum. This value increases with the DM mass. For a given DM mass, at splittings much smaller or much larger than $\delta_{\rm max}$, the evaporation rate vanishes or becomes negligible. The former case, $\delta \ll \delta_{\rm max}$, corresponds to the well-known elastic limit, where evaporation is important only for very low DM masses ($m_\chi \sim 3-4$~GeV). In the latter case, $\delta \gg \delta_{\rm max}$, evaporation becomes suppressed due to two reasons. First, halo $\chi$ particles are captured into states with, on average, lower $E$ as $\delta$ is increased, which reduces the likelihood that the particles evaporate as they subsequently transition into the lower states. Halo $\chi^*$ are captured into high $E$ states, but as these particles scatter back into $\chi^*$ in the first interaction after being captured, it is extremely likely that they drop to a significantly lower $E$ state. This inhibits their evaporation. Second, the $\chi$ scatterings may not be kinematically allowed, and if they are, the resulting $\chi^*$ end up with very little energy, and therefore in tightly bound orbits. As can be observed, the evaporation rate is extremely low over a solar lifetime, with at most $1\,(2)\,\%$ percent of particles evaporated for $m_\chi \lesssim 100\,(500)$~GeV.

\section{Summary and conclusions} \label{sec:conclusions}

In this paper we have studied the evolution of the distribution of inelastic DM in the Sun. We have presented the results of a numerical simulation of the process of DM capture and further scattering with nuclei in the Sun. We were particularly interested in the case of inelastic DM with mass splittings $\delta$ ranging from tens to hundreds of keV.

Our goal was to quantitatively study the process of thermalisation. In order for our simulation to be computationally feasible, we have neglected annihilations in the evolution equation. For definiteness, we have assumed that the DM halo consists of equal populations of the two different states.\footnote{We defer the case in which $\chi^*$ is unstable, and the case of light mediators, for future work.} We have evolved some initially captured distributions of $\chi$ and $\chi^*$ in order to study the final distributions at a time equal to the solar lifetime. We have found that $\chi^*$ are absent in the final distribution, see Fig.~\ref{fig:mx100_distevo}. We obtained a $\chi$-distribution that has not reached a stationary state at a solar lifetime, and that is far from being isothermal (Maxwell--Boltzmann) with a temperature equal to that of the solar core, see Fig.~\ref{fig:mx100_rdist}, unlike in the case of elastic scattering (c.f. Fig.~\ref{fig:mx5_distevo}). 

By assuming spherical symmetry, we have also computed an upper bound on the annihilation rate and found that it is quite suppressed for splittings larger than a few tens of keV. The exact suppression factor depends on the DM mass and the scattering rate, see Fig.~\ref{fig:anni}. 

When comparing the numerically obtained upper bound on the annihilation rate with the case in which equilibrium has taken place, it is found that equilibrium between annihilation and capture cannot be ruled out in most of the parameter space considered here (see right plot in Fig.~\ref{fig:anni}). Only for the case of DM being heavy and the mass splitting becoming large can equilibrium between the two be robustly ruled out. We have also studied evaporation and found that it plays a less important role than previously thought as it stays safely below a few percent for the splittings and DM masses considered, c.f.\ Fig.~\ref{fig:evap}.

The most phenomenologically relevant implications of this work are regarding the detection of neutrinos from DM annihilations in the Sun. The most promising cases to have a large annihilation rate are those where a non-negligible elastic cross section is also present or in the case of very small mass splittings ($\lesssim \mathcal{O}(10)$ keV).

Finally we would like to point out that it would also be interesting to pursue numerical studies of scenarios with large DM self-interactions, which would contribute to both capture and evaporation, in which case one could also incorporate DM annihilations for the same price in terms of simulation complexity.

\vspace{0.7cm}
\noindent {\bf Acknowledgements \vspace{0.1cm}\\}
We would like to thank Sergio Palomares-Ruiz for useful discussions, Simon Velander for participation during the early stages of this project, and Simon Israelsson for checks of the numerical code. We are grateful to Martin White for proof reading the revised version of the manuscript. The authors acknowledge the support from the Spanish MINECO through the ``Ram\'on y Cajal'' programme (RYC-2015-18132) and through the Centro de Excelencia Severo Ochoa Program under grant SEV-2016-0597 [M.B.], the G\"oran Gustafsson foundation~[M.B.,~S.C.], and the Australian Research Council through the ARC Centre of Excellence for Particle Physics at the Terascale (CoEPP) (CE110001104) [J.H.-G.]. S.C. also acknowledges the hospitality of Instituto de F\'isica Te\'orica (IFT) and the support from the Roland Gustafsson foundation for theoretical physics during his stay at IFT where part of this work was carried out.

%

\end{document}